\documentclass[fleqn,usenatbib]{mnras}

\usepackage[T1]{fontenc}

\usepackage{amsmath}
\usepackage{ae,aecompl}

\usepackage{booktabs}
\usepackage{microtype}

\usepackage{float}
\usepackage{caption}
\usepackage{graphicx}
\usepackage{multirow}
\usepackage{longtable}
\usepackage{tabularray}
\usepackage[inline]{enumitem}
\usepackage[dvipsnames,svgnames,table]{xcolor}

\usepackage{newtxtext}
\usepackage{newtxmath}

\usepackage{siunitx}
\usepackage{fix-cm}

\usepackage{hyperref}
\usepackage{cleveref}

\usepackage{printlen}
\usepackage[svgnames]{xcolor}

\UseTblrLibrary{booktabs, siunitx}

\graphicspath{{figs/}}

\hypersetup{
  colorlinks,
  citecolor=blue,
  linkcolor=blue,
  urlcolor=blue}

\crefname{table}{Table}{Tables}
\crefname{figure}{Figure}{Figures}
\Crefname{figure}{Fig.}{Figs.}
\crefname{equation}{Equation}{Equations}
\crefname{section}{Section}{Sections}

\setcitestyle{notesep={ },round,citesep={;},aysep={},yysep={;}}
\DeclareSIUnit{\msun}{\ensuremath{\text{M}_\odot}}
\DeclareSIUnit{\rsun}{\ensuremath{\text{R}_\odot}}
\DeclareSIUnit{\lsun}{\ensuremath{\text{L}_\odot}}
\DeclareSIUnit{\zsun}{\ensuremath{\text{Z}_\odot}}
\DeclareSIUnit{\pixel}{pixel}
\DeclareSIUnit{\pix}{pix}
\DeclareSIUnit{\dex}{dex}
\DeclareSIUnit{\arcsec}{arcsec}
\DeclareSIUnit{\arcmin}{arcmin}
\DeclareSIUnit{\yr}{yr}
\DeclareSIUnit{\year}{yr}
\DeclareSIUnit{\pc}{pc}
\DeclareSIUnit{\parsec}{pc}
\DeclareSIUnit{\Jy}{Jy}
\DeclareSIUnit{\jansky}{Jy}
\DeclareSIUnit{\sr}{sr}
\DeclareSIUnit{\steradian}{sr}
\DeclareSIUnit{\electron}{\ensuremath{\text{e}^-}}
\DeclareSIUnit{\DN}{DN}
\DeclareSIUnit{\datanumber}{DN}
\DeclareSIUnit{\abmag}{AB mag}
\DeclareSIUnit{\mag}{mag}
\DeclareSIUnit{\percentt}{per cent}
\DeclareSIUnit{\nonexistentunitjustforprefixes}{\relax}
\DeclareDocumentCommand\mast{}{\ensuremath{M_\ast}}

\DeclareDocumentCommand\ra{ m }%
  {
    \ang[angle-symbol-degree=\textsuperscript{h}, 
         angle-symbol-minute=\textsuperscript{m},
         angle-symbol-second=\textsuperscript{s}]{#1}
  }%

\author[Alejandro Guzmán-Ortega et al.]{
    \parbox{18cm}{
    Alejandro Guzmán-Ortega,$^{1}$\thanks{E-mail: a.guzman@irya.unam.mx} 
    Gustavo Bruzual,$^{1}$ 
    Vicente Rodriguez-Gomez$^{1}$, Lars Hernquist$^{2}$ \\
    }
    \vspace{0.3cm} \\
    $^{1}$ Instituto de Radioastronom\'ia y Astrof\'isica,
    Universidad Nacional Aut\'onoma de M\'exico, Apdo. Postal
    72-3, 58089 Morelia, Mexico \\
    $^{2}$ Harvard-Smithsonian Center for Astrophysics, 60 Garden Street, Cambridge, MA 02138, USA
    }
\title[Synthetic JWST galaxy images in TNG50]{Synthetic JWST galaxy images in the TNG50 simulation - I. Model validation and comparison to observations}

\begin{document}
\maketitle
\begin{abstract}
We use the TNG50 cosmological simulation and three-dimensional
radiative transfer post-processing to generate dust-aware
synthetic observations of galaxies at \( 3 \leqslant z \leqslant 6
\) and \( \log_{10} \left(\mast / \unit{\msun}\right) \geqslant
8.5 \), tailored to match the depth and resolution of current deep
JWST surveys (NGDEEP and JADES). We analyse the performance of
spectral energy distribution (SED) fitting on the simulated
sample, focusing on the recovery of photometric redshift and
stellar mass. At \( z \leqslant 5 \), we find that
\qty{>90}{\percentt} of redshifts are recovered within
\num{\pm0.2}, but performance declines at \( z = 6 \). Stellar
masses are generally well-recovered within a factor of \num{2},
but are systematically underestimated regardless of redshift, a
trend that is more pronounced at the high-mass end \(
\left(\log_{10}\left(\mast / \unit{\msun}\right) \geqslant
10\right) \). In addition, we study the observer-frame colours of
galaxies in this redshift range as well as the SED-inferred
\emph{UVJ} diagram. We find that TNG50 galaxies broadly follow the
tendencies marked by observations, but tend to be slightly redder
at lower masses and bluer at higher masses, regardless of
redshift. Finally, using a colour-based definition of quiescence,
we determine the fraction of quiescent galaxies as a function of
stellar mass at \( 3 \leqslant z \leqslant 6 \), which we find to
be broadly consistent with observations.
\end{abstract}

\begin{keywords}
    methods: numerical - techniques: image processing - galaxies: formation - galaxies: photometry
\end{keywords}

\section{Introduction} 
    \label{sec:introduction}

Cosmological simulations have become an indispensable resource for
studying the large-scale structure of the Universe as well as the
formation and evolution of galaxies. These simulations have
attained a high level of sophistication, not only in terms of the
physical processes they include but also in the resolution they
achieve, which allows for detailed studies of both individual
galaxies and statistically representative samples \citep[for
recent reviews, see][]{Vogelsberger2020,Crain2023}. In particular,
large-box simulations like IllustrisTNG
\citep[][]{Pillepich2018,Nelson2019a}, Horizon-AGN
\citep[][]{Dubois2014}, EAGLE \citep[][]{Schaye2015,Crain2015},
MUFASA \citep[][]{Dave2016}, and SIMBA \citep[][]{Dave2019} have
been able to broadly reproduce several key properties of the real
galaxy population, including, to name a few, the colour bimodality
\citep[e.g.][]{Trayford2017,Nelson2019,Cui2021}, the star-forming
main sequence \citep[e.g.][]{Donnari2019,Matthee2019}, the
mass--size relation
\citep[e.g.][]{Furlong2017,Genel2018,Dave2019}, the
mass--metallicity relation
\citep[e.g.][]{DeRossi2017,Torrey2019,Dave2019} and their visual
morphologies
\citep[][]{Rodriguez-Gomez2019,Huertas-Company2019,Bignone2020,Guzman-Ortega2023}.
This agreement with observations has positioned these simulations
as powerful tools for interpreting the Universe and to predict the
evolution of galaxies across cosmic time.

Despite this, conducting direct comparisons between simulations
and observations is not always straightforward, largely due to the
need for a consistent link between these two domains. Usually,
this is addressed through
\begin{enumerate*}[leftmargin=*, label=(\emph{\roman*})]
    \item \emph{inverse modelling}, where physical properties are
    extracted from observations to compare against simulation
    outputs, or
    \item \emph{forward modelling}, where synthetic observations
    are generated from simulations and then compared to actual
    observational data.
\end{enumerate*}
Regarding the latter approach, forward modelling has found
numerous applications, to name a few, these include: determining
galaxy merger time-scales via synthetic images of merging galaxies
\citep[e.g.][]{Lotz2008a,Lotz2010}; to study dust scaling
relations and their agreement with observations
\citep[e.g.][]{Camps2016,Kapoor2021}; to determine dust
attenuation and emission in high-redshift galaxies
\citep[e.g.][]{Ma2019,Cochrane2019,Schulz2020}; to provide
expectations for luminosity functions at low and high redshift
\citep[e.g.][]{Vogelsberger2020,Trcka2022}; and to study the
morphology and sizes of galaxies at low and high redshift
\citep[e.g.][]{Rodriguez-Gomez2019,Bignone2020,Parsotan2020,Guzman-Ortega2023,Costantin2023}.

Apart from the previous tasks, another compelling application of
forward modelling is to evaluate the performance of spectral
energy distribution (SED) fitting techniques \citep[see
e.g.][]{Wuyts2009,Torrey2015,Hayward2015,Smith2018,Katsianis2020}.
These methods are widely adopted to estimate the physical
properties of galaxies using broadband photometry, spectral data,
or a combination of both. They have become a standard tool in
observational astronomy, particularly in the study of galaxy
evolution. Coupling forward modelling with SED fitting provides
the advantage of having ground truth values for most of the
properties of interest, an aspect that is often missing with real
data and that hinders the validation of SED fitting methods.
Despite this, the reliability of these methods is often
questioned, as they depend on several assumptions that can
introduce biases and uncertainties in the derived properties. To
name a few, these include the choice of stellar population
synthesis (SPS) models, the treatment of dust, the assumed form of
the star formation history (SFH), and the presence of active
galactic nuclei
{\citep[e.g.][]{Walcher2010,Conroy2013,Jones2022a,Pacifici2023,Bellstedt2025}}.
For recovering stellar masses, several studies have highlighted
the impact of these assumptions: \citet{Michalowski2012} found
that the parametric form of SFHs can significantly affect the
estimated stellar masses of submillimetre galaxies. Similarly,
\citet{Mitchell2013} showed that simplified dust screen models can
lead to substantial underestimations of stellar masses,
particularly for galaxies with high dust content. In addition,
\citet{Lower2020}, using synthetic galaxy spectra at \( z \sim 0
\) from the SIMBA simulation, demonstrated that non-parametric
SFHs can significantly improve stellar mass recovery compared to
parametric SFHs. {{Using synthetic EAGLE spectra at \( 1 < z < 4
\), \citet{Katsianis2020} showed that the tension in the star
formation rate to stellar mass relation between simulations and
observations can be largely reduced when consistent methods are
used to infer physical properties.}} {{In a similar vein,
\citet{Faucher2024} fit SEDs to simulation-derived photometry and
determined that mismatches between models and true galaxy
properties (e.g. the form of the SFH) can lead to a systematic
underestimation of recovered parameters, biasing this same
relation.}} More recently, \citet{Cochrane2025}, employing a
similar approach for synthetic spectra at \( z \geqslant 7 \),
emphasized that poorly accounting for line emission during fitting
can introduce significant offsets in the derived stellar masses.
These findings highlight the importance of considering these
factors when interpreting the results of SED fitting.

Another important application of synthetic observations is the
comparison of simulated and observed galaxy colours using
diagnostic diagrams such as the \emph{UVJ} diagram
{{\citep[e.g.][]{Williams2009,Whitaker2011,Muzzin2013}}}. This
diagram serves as a valuable tool for distinguishing and analysing
the properties of star-forming galaxies (typically located in the
bottom-left corner), quiescent galaxies (upper-left region), and
dusty star-forming galaxies (upper-right region). Indeed, it is
often used to test whether current simulations of galaxy formation
can reproduce the observed colour distributions of galaxies. At \(
z \lesssim 1-2 \), this has been moderately achieved by several
studies: \citet{Trayford2017} forward-modelled a sample of
galaxies from the EAGLE simulation at \( z = 0.1 \), broadly
reproducing the basic features of the observed \emph{UVJ} diagram
but finding an excess of star-forming galaxies masquerading as
quiescent; \citet{Dave2017} created dust-corrected synthetic
photometry for galaxies from the MUFASA simulation and found that
quiescent and star-forming galaxies are well separated with
traditional selection criteria; \citet{Donnari2019} followed a
similar approach with the TNG100 and TNG300 simulations, and
reached comparable results; \citet{Gebek2024} compared the
observational \emph{UVJ} diagram with its counterpart from the
TNG100 simulation at \( z \leqslant 0.1 \), finding good agreement
between them, but noting that the simulated sample presented a
systematic reddening in both colour axes relative to observations. 

However, at \( z \gtrsim 2 \), reproducing the observed \emph{UVJ}
diagram has proved challenging. Observations indicate that the
upper-right region of the diagram is populated by massive (\(
\log_{10} \left( \mast / \unit{\msun} \right) \gtrsim 10.5 \))
dusty star-forming galaxies \citep[e.g.][]{Skelton2014,Fang2018},
a feature that simulations struggle to reproduce. {{At \( z \sim 2
\), \citet{Dave2017}, \citet{Donnari2019}, and more recently
\citet{Gebek2025} all found that, despite being in broad agreement
with observations, their corresponding diagrams failed to
reproduce the observational range of colours (showing redder \( U
- V \) but bluer \( V - J \)) and a deficit of massive dusty
star-forming galaxies.}} Likewise, \citet{Akins2022} created mock
observations at \( z = 1\text{--}2 \) from the SIMBA simulation
with its on-the-fly dust model, and found that, while the
clustering of quiescent and star-forming galaxies was well
reproduced, the upper-right region was populated by low-mass
galaxies, in tension with observational trends.

In this work, we describe the generation of synthetic observations
of high-redshift galaxies from the TNG50 simulation, including
radiative transfer calculations to account for dust effects, which
are subsequently post-processed to match the depth and resolution
of deep JWST surveys. We then carry out photometric measurements
on these synthetic images in order to explore two main topics.
Firstly, we explore the accuracy of SED fitting tools on TNG50
galaxies at \( 3 \leqslant z \leqslant 6 \), with emphasis on the
recovery of photometric redshift and stellar mass. Secondly, we
examine the colours of TNG50 galaxies in this redshift range, with
particular interest in the \emph{UVJ} diagram, and compare them
against observations from current JWST surveys.

This paper is structured as follows. \cref{sec:data} describes the
TNG50 simulation and the observational datasets used in this work.
In \cref{sec:methods}, we present our methodology for generating
dust-aware JWST-like synthetic observations of high-redshift
galaxies, and describe the SED fitting for estimating both
photometric redshifts and stellar masses. We present the results
of our analysis, including the performance of SED fitting, both
the observer-frame and the estimated \emph{UVJ} colours of
galaxies, as well as the mass-dependent quiescent fraction in the
\( 3 \leqslant z \leqslant 6 \) range, and how they compare to
current observations in \cref{sec:results}. We discuss our
findings in \cref{sec:discussion} and present our conclusions in
\cref{sec:conclusions}. We assume a flat \(\Lambda\)CDM cosmology
consistent with IllustrisTNG, following
\citet{PlanckCollaboration2016}, and use the AB magnitude system
\citep{Oke1983}.

\section{Data} 
    \label{sec:data}

\subsection{The TNG50 simulation}
    \label{subsec:illustristng} 

The TNG50 simulation \citep{Pillepich2019,Nelson2019a} is the
highest-resolution run of the IllustrisTNG project
\citep{Marinacci2018,Naiman2018,Nelson2018,Pillepich2018,Springel2018},
a suite of cosmological magnetohydrodynamical simulations of
galaxy formation and evolution. The TNG50 simulation follows the
evolution of a periodic box of side length
\qty{51.7}{\mega\parsec} from redshift \( z = 127 \) to \( z = 0
\), with a dark matter particle resolution of \qty{4.5e5}{\msun}
and a baryonic particle resolution of \qty{8.5e4}{\msun}. The
galaxy formation model employed in the IllustrisTNG project
includes a comprehensive treatment of several key physical
processes, such as gas cooling and heating, star formation and
evolution, metal enrichment, black hole growth, and feedback from
active galactic nuclei and supernovae. Dark matter halos are
identified via friends-of-friends groups using the percolation
algorithm of \citet{Davis1985}, and subhalos are found using the
\textsc{subfind} algorithm \citep{Springel2001,Dolag2009}. For a
detailed description of the simulation setup and the galaxy
formation model, we refer the reader to \citet{Pillepich2019} and
{\citet{Nelson2019a}}.

\subsection{The Next Generation Deep Extragalactic Public (NGDEEP)
    Survey} 
    \label{subsec:ngdeep} 

The NGDEEP survey is a deep slitless spectroscopic and imaging
survey JWST Cycle 1 treasury program \citep[PID 2079; PIs: S.
Finkelstein, C. Papovich, N. Pirzkal]{Bagley2024} aimed to study
feedback processes in galaxies at \( 1 \lesssim z \lesssim 12 \).
In its spectroscopic mode, the survey targets the HUDF and
simultaneously images the HUDF05-02 parallel field. For the
latter, the survey provides deep NIRCam imaging in the F115W,
F150W, F200W, F277W, F356W and F444W filters (\qtyrange{\sim
1}{5}{\micro\meter}) totalling a combined exposure time of
\qty{\sim 190}{\hour}, with \( 5\sigma \) depths of
{{\qtylist{30.7;30.8}{\abmag} for F444W and F356W,
\qty{30.9}{\abmag} for F150W, F200W and F277W, and
\qty{31.2}{\abmag} for F115W}}. This program was initially
scheduled for completion in early 2023. However, due to a
temporary suspension of JWST operations, only the first half was
completed at that time, with the second half finalized {{in
2025}}. In this work, we use NIRCam imaging mosaics and
photometric catalogues from both rounds of observations, as
presented in \citet{Leung2023} and Leung et al. (in prep.). We
refer the reader to these works and references therein for a
detailed description of the data reduction and photometric
measurements.

\subsection{The JWST Advanced Deep Extragalactic Survey (JADES) program}
    \label{subsec:jades}

The JADES program is a JWST Cycle 1 guaranteed time observations
program \citep[PIDs 1180, 1181, 1210, 1286; PIs: D. Eisenstein, N.
Luetzgendorf]{Eisenstein2023} aimed at investigating galaxy
formation from high redshift to cosmic noon. It targets the
GOODS-S and GOODS-N fields, providing infrared imaging and
spectroscopy with NIRCam, NIRSpec, and MIRI. In the GOODS-S field,
it provides deep imaging over an area of approximately
\qty{45}{\arcmin\squared} using nine NIRCam filters, with a total
exposure time of around \qty{130}{\hour}. In this work, we utilize
both imaging data and photometric catalogues from the initial
release of the GOODS-S field as presented in \citet{Rieke2023}.
This dataset covers a smaller area of about
\qty{25}{\arcmin\squared} with an exposure time of approximately
\qty{111}{\hour}, achieving \( 5\sigma \) depths of
{{\qtylist{29.82;29.74;29.79;30.07;30.01;29.77}{\abmag} for F115W,
F150W, F200W, F277W, F356W and F444W, respectively. We refer to
\citet{Rieke2023} for a description of the data and reduction
process.}}

\begin{table}
    \centering
    \caption{Summary of the TNG50 galaxy sample used in this work. We consider \num{13} snapshots from redshift \( z = 6 \) to \( z = 3 \), imposing a stellar mass cut of \( \log_{10}\left( \mast / \unit{\msun} \right) \geqslant 8.5 \). The table shows the cumulative number of galaxies at each redshift, as well as the median and maximum stellar masses. The column showing the median values also indicates the 16th and 84th percentiles as subscripts and superscripts, respectively.}
    \label{tab:tng50sample}
    \begin{tblr}{
        width=\columnwidth,
                    colspec={
                    Q[si={table-format=1.2},l,co=1]
                    Q[si={table-format=5},c,co=1]
                    Q[si={table-format=3.2(2)},c,co=1]
                    Q[si={table-format=2.2},r,co=1]
                    },
                    row{1,2}={guard},
                    row{1}={belowsep=0pt}
                    }
                    \toprule
                    Redshift & {Cumulative \\ sum} & {Median \\ stellar mass} & {Maximum \\ stellar mass} \\
                    \( z \) & & \( \log_{10} \left( \mast /
                    \unit{\msun} \right) \) & \( \log_{10} \left( \mast /
                    \unit{\msun} \right) \) \\
                    \midrule
                    6 & 161 & 8.72(0.35:0.18) & 10.61 \\
                    5.8 & 357 & 8.71(0.42:0.16) & 10.69 \\
                    5.5 & 626 & 8.72(0.45:0.16) & 10.83 \\
                    5.2 & 998 & 8.73(0.44:0.18) & 10.89 \\
                    5 & 1462 & 8.75(0.43:0.19) & 10.96 \\
                    4.7 & 2083 & 8.77(0.47:0.20) & 11.04 \\
                    4.4 & 2862 & 8.77(0.49:0.19) & 11.10 \\
                    4.2 & 3786 & 8.81(0.50:0.21) & 11.15 \\
                    4 & 4830 & 8.82(0.55:0.23) & 11.24 \\
                    3.7 & 6141 & 8.85(0.57:0.26) & 11.33 \\
                    3.5 & 7660 & 8.87(0.59:0.28) & 11.39 \\
                    3.3 & 9409 & 8.87(0.63:0.28) & 11.44 \\
                    3 & 11502 & 8.86(0.69:0.27) & 11.48 \\
    \bottomrule
    \end{tblr}
\end{table}

\subsection{Galaxy sample} 
    \label{subsubsec:galaxysample} 

The simulated galaxy sampled used in this work consists of all
TNG50 subhalos at \( 3 \leqslant z \leqslant 6 \) with stellar
masses \( \log_{10}\left( \mast / \unit{\msun} \right) \geqslant
8.5 \), totalling \num{11502} objects. This is summarized in
\cref{tab:tng50sample}. Similarly, the observational galaxy sample
consists of all NGDEEP and JADES objects whose photometric
redshifts fall within the same range. These redshifts were
estimated using the \texttt{EAZY} code \citep{Brammer2008} as part
of the data reduction and photometric estimations presented in
\citet{Leung2023} and \citet{Rieke2023} for NGDEEP and JADES,
respectively. The joint observational sample contains {{\num{823}
galaxies, \num{230} from NGDEEP and \num{593} from JADES,
satisfying the same selection criteria used for the simulated
sample. We note there is no overlap between the subsamples}}.
Stellar masses for the observational datasets were estimated with
a commonly-used SED fitting code, as described in
\cref{subsubsec:obssedfitting}. {{We note that
\citet{Simmonds2024} have estimated, for the full JADES sample, a
\qty{90}{\percentt} mass completeness limit of \(
\log_{10}\left(\mast/\unit{\msun}\right) \approx 7.5 \) at \( z
\sim 3-9 \), which supports our choice of stellar mass cut for
both samples.}}

\section{Methods} 
    \label{sec:methods}

\subsection{Synthetic observations}
    \label{subsec:syntheticobservations}

In this section we outline the creation of NGDEEP-like synthetic
observations of TNG50 galaxies. We first describe the process to
generate synthetic images including full radiative transfer, and
then we present the method to include realism to match real JWST
observations.

\subsubsection{Radiative transfer with SKIRT}
    \label{subsubsec:skirt}

The \texttt{\textsc{SKIRT}} code \citep{Camps2015,Camps2020} is a
Monte Carlo radiative transfer code for primarily simulating dust
effects on the radiation of astrophysical systems. During a
simulation run, the radiation field from sources is discretized
into photon packets which traverse a gridded domain of a dusty
medium. The path of each packet is traced as it is absorbed,
scattered or re-emitted by the dust in the system; the packets are
then captured as they escape the system by recording instruments
that can output a spectral energy distribution (SED) or a spectral
datacube. Since the code adopts a Monte Carlo approach, it
supports arbitrary geometries and dust distributions, which makes
it suitable for post-processing objects from hydrodynamical
simulations. We follow the procedure outlined in
\citet{Guzman-Ortega2023}, which was based on earlier works by
\citet{Camps2016}, \citet{Trayford2017} and
\citet{Rodriguez-Gomez2019}.

\subsubsection{Primary emission sources}
    \label{subsubsec:primarysources}

{{Stellar particles are treated as simple stellar populations and
used as primary emission sources for a \texttt{SKIRT} simulation.
For a given subhalo, we extract all stellar particles belonging to
the same FoF group within a}} spherical aperture that is centred
on the object's most-bound particle, and  whose radius is equal to
\( \max ( 7.5\, r_{1/2,\,\ast},\, r_{1/2,\,\text{tot}} ) \), where
\( r_{1/2,\,\text{tot}} \) is the radius that contains half of the
total mass of the subhalo and \( r_{1/2,\,\ast} \) is the stellar
half-mass radius. In addition, we compute for each particle a
smoothing scale given as the distance to its 32nd nearest
neighbour, used by \texttt{SKIRT} to compute a smooth particle
distribution.

Particles whose age is greater than \qty{10}{\mega\year} are
assigned an SED from the revised version of the
\citet[][BC03]{Bruzual2003} stellar population synthesis models
introduced by \citet[][CB19 models hereafter]{Plat2019}. The CB19
models represent a major revision of the BC03 models, and follow
the PARSEC evolutionary tracks \citep[][]{Marigo2013,Chen2015} for
16 different chemical compositions \citep[listed in table 8
of][]{Sanchez2022}. These evolutionary tracks describe the
evolution of stars with masses from \qtyrange{0.1}{600}{\msun},
using a fine grid of mass and time steps. The tracks run from the
main sequence to the Wolf-Rayet phase for massive stars and up to
the thermally pulsing asymptotic giant branch (TP-AGB) for stars
below \qty{6}{\msun}. In the CB19 models, the evolution of the
post-AGB phase of intermediate- and low-mass stars follows the
prescription of \citet[]{MillerBertolami2016}. The PARSEC tracks
assume that the initial solar nebula had \( Z = 0.014 \) and the
current Sun has a surface abundance of \( Z = 0.017 \). Tables 9
to 12 of \cite{Sanchez2022} describe in detail the stellar
spectral libraries used to build the CB19 models. In what follows,
we use the CB19 models computed for the \citet{Chabrier2003}
initial mass function with an upper mass limit of
\qty{100}{\msun}. 

{{The colour evolution of the CB19 and BC03 models is formally
similar when using evolutionary tracks and stellar libraries of
the same metallicity. The finer mass grid used in the PARSEC
tracks, as compared to the coarser mass grid in the PADOVA 1994
\citep[][]{Alongi1993,Bressan1993,Fagotto1994,Fagotto1994a,Girardi1996}
tracks used by BC03, translates into a smoother spectral and
colour evolution in the CB19 than in the BC03 models. This is
particularly evident when comparing the CB19 models built with the
MILES and INDOUS stellar libraries
\citep[][]{Sanchez-Blazquez2006,Falcon-Barroso2011,Prugniel2011,Valdes2004}
to the BC03 models, which use the lower-resolution and less
complete STELIB library \citep[][]{Borgne2003}. For SED fitting
and SFH reconstruction, the CB19 models, available for 16
different metallicities, provide in general smoother solutions
than the BC03 models, available only for 6 different
metallicities. Both set of models suffer from the age-metallicity
degeneracy \citep[][]{Worthey1994} and when only a rough estimate
of the age and metallicity of a stellar population is required,
the CB19 and BC03 solutions are comparable.}}

Similarly, stellar particles whose age is less than
\qty{10}{\mega\year} are assumed to be in star-bursting regions and
embedded in dust. As such, their SEDs are set using models from
the \texttt{MAPPINGS-III} code \citep{Groves2008}, which include
emission from {H \textsc{ii}} regions, photo-dissociation regions,
as well as absorption by gas and dust present in the birth clouds.
These templates are parametrized by
\begin{enumerate*}[leftmargin=*, label=(\emph{\roman*})]
    \item the star formation rate (SFR),
    \item the metallicity \( Z \), 
    \item the interstellar medium (ISM) pressure \( P_0 \),
    \item the compactness \( \mathcal{C} \) of the region, and
    \item the cloud covering fraction \( f_\text{PDR} \).
\end{enumerate*}
The first two parameters are obtained directly from the simulation
data. For the third and fourth parameters we assume fixed
\emph{typical} values of \( P_0 = \qty{1.38e-12}{\pascal} \) and
\( \log_{10} \mathcal{C} = 5 \) \citep[see
e.g.][]{Rodriguez-Gomez2019,Schulz2020}. Lastly, we set \(
f_\text{PDR} = 0.2 \) following \citet{Jonsson2010}, noting that
other works \citep[e.g.][]{Schulz2020} have found no significant
impact {{when}} changing this value.

\subsubsection{Dust distribution}
    \label{subsubsec:dustdistribution}

Since the TNG50 simulation does not explicitly track ISM dust, we
infer its distribution using the gas properties from the
simulation. We assume that the diffuse dust component of a given
subhalo is traced by its gas-phase metal content, also imposing
the condition that gas cells must be star-forming (i.e. with
density above the star formation threshold). The exact amount of
metals that is locked into dust grains is set by the
dust-to-metals ratio, \( f_\text{dust} \), which is often treated
as a free parameter in most modelling works. Although there are
indications, {{from both simulations and observations}}, that this
quantity could be a function of redshift
\citep[e.g.][]{Vogelsberger2020} or gas-phase metallicity
{\citep[e.g.][]{Remy-Ruyer2014,Popping2017}} the exact trend and
dependency on these parameters remains uncertain, thus we assume a
constant value of \( f_\text{dust} = 0.3 \) remaining consistent
with previous work
\citep[e.g.][]{Rodriguez-Gomez2019,Guzman-Ortega2023}.

The dust distribution of a subhalo is therefore obtained from its
gas content following the expression
\begin{equation}
    \rho_\text{dust} = \begin{cases}
        f_\text{dust} \, Z \, \rho_\text{gas} & \text{if } \rho_\text{gas} > \rho_\text{thr} \\
        0 & \text{otherwise}
    \end{cases},\label{eq:dustdistribution}
\end{equation}
where \( \rho_\text{dust} \) is the dust density, \( Z \) is the
gas metallicity, and \( \rho_\text{thr} \) is the star formation
density threshold (\( n_\text{H} =
\qty{0.13}{\per\centi\meter\cubed} \)). 

We note that there is no unique way of defining the allocation of
dust. For example, other studies
\citep[e.g.][]{Camps2016,Trayford2017} have used a modified version
of \cref{eq:dustdistribution}, imposing a maximum temperature, so
that dust can also be present if the gas cell temperature is below
such threshold. Since most simulations do not track dust directly,
this maximum temperature is difficult to constrain and the
threshold is chosen somewhat arbitrarily. Another approach
consists in separating the hot circumgalactic medium from the
cooler gas, adopting a line cut in the temperature-density plane
\citep{Torrey2012}. In any case, the star-forming gas mainly
traces its true dust content, so that assuming different recipes
for assigning this distribution should not have a significant
impact on the results as long as this assertion is met
\citep[e.g.][]{Kapoor2021}.

The dust composition associated with this distribution is assumed
to follow the \citet{Zubko2004} dust model. This model includes a
mixture of graphite, silicate grains, and polycyclic aromatic
hydrocarbons. This composition has been found to be consistent
with various observational constraints, including extinction
curves, infrared emission, and elemental abundances. Regarding the
spatial grid for the dusty medium, we adopt a Voronoi grid so that
it matches the one used in TNG50, and that is uniquely constructed
from the gas cell positions, which represent the mesh generating
points. {{This grid has cubical geometry with a side length equal
to twice the aperture radius used to extract the stellar
particles.}}

\subsubsection{Other parameter settings}
    \label{subsubsubsec:othersettings}

For each \texttt{SKIRT} simulation, we define frame instruments to
record broadband images. The field of view is set to twice the
aperture radius used to extract the stellar particles. The
resulting images are in the observer's frame, accounting for
cosmological surface brightness dimming at each redshift. The
wavelength coverage corresponds to the NIRCam filters F115W,
F150W, F200W, F277W, F356W, and F444W, spanning approximately
\qtyrange{0.6}{5.1}{\micro\meter}. The pixel scale is set to
\ang[angle-symbol-over-decimal]{;;0.031} \unit{\per\pixel} for the
first three filters and \ang[angle-symbol-over-decimal]{;;0.063}
\unit{\per\pixel} for the remaining ones. Additionally, we set the
number of photon packets to \num{3e8}, which we find sufficient
for our purposes. Lastly, we set three different viewing angles
for the frame instruments, which correspond to the \emph{xy},
\emph{xz}, and \emph{yz} spatial projections, which yields a total
of \( \numproduct{3 x 6 x 11502} = \num[evaluate-expression]{3 * 6
* 11502}\) unique images for the sample.

\begin{figure*}
    \centering
    \includegraphics{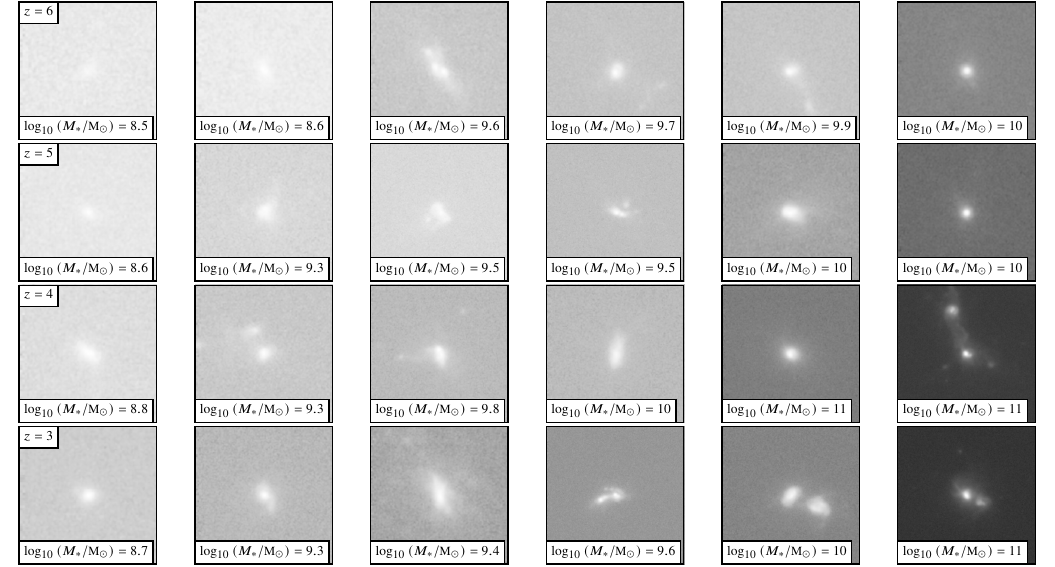}
    \includegraphics{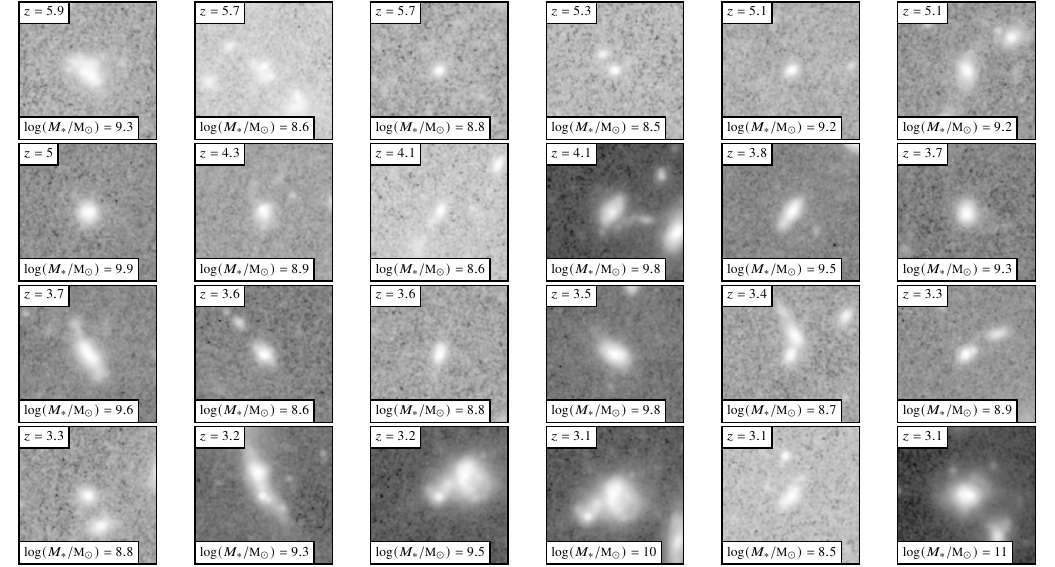}
    \caption{\textbf{First four rows:} example TNG50 NGDEEP-like
    synthetic images in F356W at \( z=3, 4, 5\) and \( 6 \). Each
    row shows galaxies at the same redshift, with objects sorted
    from left to right by increasing stellar mass
    (\textsc{subfind} values). We only include close companions in
    the same host halo as the central object. \textbf{Last four
    rows:} example NGDEEP galaxies in F356W, ordered from top to
    bottom by decreasing photometric redshift. As opposed to the
    synthetic images, these contain a contribution from foreground
    and background objects that, in most cases, are not physically
    associated with the main galaxy.}
    \label{fig:synth-survey_panels}
\end{figure*}

\begin{table} \sisetup{round-mode=places, round-precision=3,
             }
    \centering
    \caption{Mean background levels in units of
    \unit{\mega\jansky\per\steradian} for each NIRCam filter used
    in the synthetic observations. These values were measured from
    the corresponding NGDEEP mosaics.}
    \label{tab:bkg_lvls}
    \begin{tblr}{width=\linewidth,
        colspec={
        X[l]
        Q[si={table-format=1.3},c]
        X[l]
        Q[si={table-format=1.3},c]
        },
        row{1}={guard},
        rowsep=0pt,
        row{1}={belowsep=0pt, abovesep=0pt}
        }
        \toprule
        {Filter} & {Mean \\ background level} &
        {Filter} & {Mean \\ background level} \\
        \midrule
        F115W & 0.1987910784477508 & F277W & 0.09757662273254215 \\
        F150W & 0.1703176584563132 & F356W & 0.10051845588732898 \\
        F200W & 0.1442664046197524 & F444W & 0.30277013816680826 \\
        \bottomrule
    \end{tblr}
\end{table}

\subsubsection{Observational realism}
    \label{subsubsec:observationalrealism}

The images generated by \texttt{SKIRT} simulations are idealized
representations of galaxies, without instrumental and
observational effects. To make them comparable to real
observations, we apply post-processing steps to simulate these
effects. 

We first convolve the images with a point spread function (PSF)
model specific to each filter. These PSF models were obtained from
the \texttt{webbpsf} package \citep{Perrin2014}, which provides
representations of the expected PSF for JWST instruments and
filters. We then create three dithered versions of each image
matching the dither pattern used in NGDEEP, and add the mean
background level (at each filter) present in NGDEEP mosaics.
Subsequently, we generate Poisson noise on the electron count in
each pixel, and use the \texttt{drizzle} package to combine the
images into a single drizzled image at the pixel scale of
\ang[angle-symbol-over-decimal]{;;0.03} \unit{\per\pixel}, which
is the same as the NGDEEP mosaics. The mean background level is
{{then}} subtracted from the image, and the pixel values are
expressed in surface brightness units of
\unit{\mega\jansky\per\steradian}. \cref{tab:bkg_lvls} lists the
mean background levels for each NIRCam filter used in these steps.
This process yields background-subtracted images that are
comparable to reduced NGDEEP data products. {{This procedure
differs from the one in \citet{Guzman-Ortega2023}, where Gaussian
background noise was added separately, and is more similar to the
one recently adopted by \citet{Sazonova2024} and
\citet{LaChance2025}, and also by \citet{Zhou2025} when dark and
read-out noise are negligible compared to the sky background noise
\citep[see][for alternative methods using survey limiting
magnitudes]{Marshall2022,Merlin2023}.}}

In \cref{fig:synth-survey_panels}, the first four rows show
examples of realistic images of TNG50 galaxies in the F356W
filter, while the remaining rows show objects from the NGDEEP
survey. For the TNG50 galaxies, each row displays galaxies at the
same redshift, with objects sorted from left to right by
increasing stellar mass. {{Note that neighbouring galaxies from
the same FoF group are included in the images by construction, and
are subsequently removed from our analysis (masked) as described
in \cref{subsubsec:syntheticphotometry}.}} These images are
representative of the full sample considered in this work.

\begin{table*}
    \centering
    \caption{Free parameters and their corresponding priors used
    in the SED fitting procedure for the simulated and
    observational samples. Logarithmic priors are applied in base
    ten. We note that the redshift is fixed to the true value for
    the simulated sample, while it is left as a free parameter for
    the observational samples. In the latter case, we set a
    Gaussian prior centred on the estimated photometric redshift
    with a standard deviation of \num{0.25}.}
    \label{tab:sedfittingparams}
    \sisetup{
  exponent-mode = threshold,
  exponent-thresholds = -4:4,
  range-exponents = individual,
  parse-numbers = false,
  input-decimal-markers={,},
  output-decimal-marker = \text{~to~},
  }
    \begin{tblr}{
        width=\textwidth,
        colspec={
        X[l]
        X[c]
        X[c]
        Q[si={},c]
        X[r]
        },
        rowsep=0pt,
        row{1} = {guard, belowsep=0pt, abovesep=0pt},
        row{4,6,9} = {belowsep+=1pt},
        cell{2}{4}={guard}
    }
    \toprule
    Component & Parameter & Symbol (unit) & Range & Prior \\
    \midrule
    \SetCell[r=3]{m} General & Redshift & \( z \) & \( z_\text{phot}\pm0.75 \) & Gaussian \\ & Total stellar mass formed & \( \mast \,
    \left( \unit{\msun} \right) \) & 1,10^{13} & Logarithmic \\
    & Stellar and gas metallicity & \( Z \, \left( \unit{\zsun} \right) \)
    & 0.001,2.5 & Uniform \\
    \SetCell[r=2]{m} Star formation history & Star formation peak
    time & \( t_\text{max} \, \left( \unit{\giga\year} \right) \)
    & 0.001,15.0 & Uniform \\
    & FWHM of SFH & \( \sigma_\text{SFH} \, \left( \unit{\giga\year}
    \right) \) & 0.001,20.0 & Uniform \\
    \SetCell[r=3]{m} Dust attenuation & Power-law index & \(
    \delta \) & -1.6,0.4 & Uniform \\
    & \qty{2175}{\angstrom} bump strength & \( B \) & 0.0,10.0 &
    Uniform \\
    & \( V \)-band attenuation & \( A_V \, \left( \unit{\mag}
    \right) \) & 0.01,8.0 & Uniform \\
    & Birth cloud obscuration & \( \eta \) & 1.0,3.0 &
    Uniform \\
    Nebular emission & Ionization parameter &
    \( \log_{10} U \) & -4.0,-2.0 & Uniform \\
    \bottomrule
    \end{tblr}
\end{table*}

\subsection{Spectral energy distribution (SED) fitting}
    \label{subsec:sedfitting}

In this section, we first outline the procedure used to measure
synthetic photometry from the simulated images, focusing on the
subsamples at \( z = 3, 4, 5, 6 \). We then describe the
estimation of photometric redshifts and recovery of physical
parameters using commonly adopted SED fitting codes. Finally, we
describe equivalent methods applied to the observational samples
from NGDEEP and JADES. 

\subsubsection{Measuring synthetic photometry}
    \label{subsubsec:syntheticphotometry}

To measure synthetic photometry from the realistic images, we use
the \texttt{photutils} package \citep{Bradley2024}. The process
begins by creating  a detection image in the F200W filter. As a
first step, we convolve the image with a \( \numproduct{7x7} \)
Gaussian kernel with a full width at half maximum (FWHM) of
\qty{4}{\pixel}. We then identify sources by using the
\texttt{detect\_sources} routine with a minimum threshold of
\num{0.75} and a requirement of at least \num{5} connected pixels.
This is followed by a deblending step, using \num{16} deblending
levels and a contrast parameter of \num{0.001}. Once the
segmentation map is defined for each object, we create a source
catalogue that includes the centroids of each source as well as
the Gaussian-equivalent semimajor and semiminor elliptical axes,
the angular orientation and the Kron radius. The Kron radius is
computed setting the Kron parameter to \( K=1.5 \). The photometry
is then measured across all filters using the derived Kron
apertures and the pixel flux sum within each object's segmentation
map. Uncertainties on these measurements are computed by combining
background noise with the Poisson noise of the sources. The total
error, \( \sigma_\text{tot} \), is given by \( \sigma_\text{tot} =
\frac{1}{g_\text{eff}} \sqrt{g^2_\text{eff} \sigma^2_\text{bkg} +
g_\text{eff} I} \), where \( \sigma_\text{bkg} \) is the
background noise, \( g_\text{eff} \) is the effective gain (i.e.
the ratio of electrons to image units), and \( I \) corresponds to
the image data in units of \unit{\mega\jansky\per\steradian}. From
this point on, we only consider measurements derived from the
\emph{xy} projection of the images and whose final size is at
least \qty{32}{\pixel}.

\subsubsection{Estimating photometric redshifts for the simulated sample}
    \label{subsubsec:synthphotz}

We estimate photometric redshifts for individual subsamples at the
fixed redshifts \( z = 3, 4, 5, 6 \) using the \texttt{EAZY} code
\citep{Brammer2008}. This code fits non-negative linear
combinations of a pre-generated template set to the input
photometry by minimizing the \( \chi^2 \) statistic between the
input and template fluxes. We use the default set of templates
(named \texttt{tweak\_fsps\_QSF\_12\_v3} in the code) for all
subsamples. We disable priors on magnitude and the UV slope, and
do not perform iterative zero-point offset estimation.
Additionally, we set an error floor on the photometry of
\qty{5}{\percent} to account for potential differences between the
templates and our sample. Lastly, the maximum likelihood redshift,
corresponding to the peak of the redshift probability
distribution, is taken as the best estimate.

\subsubsection{Estimating physical parameters for the simulated sample}
    \label{subsubsec:synthsedfitting}

We estimate the physical properties of the simulated sample using
the \texttt{BAGPIPES} code \citep{Carnall2018}. This code employs
a Bayesian approach to generate model spectra of galaxies spanning
the ultraviolet to the microwave regimes, and to fit these models
to observational data using the \texttt{MULTINEST} nested sampling
algorithm \citep[][]{Feroz2009}. The fitting we perform to the
synthetic photometry includes the following components:
\begin{enumerate*}[leftmargin=*, label=(\emph{\roman*})]
    \item SPS templates from the 2016
    version \citep[][]{Chevallard2016} of the \citet{Bruzual2003}
    models;
    \item line and nebular continuum emission through the
    ionization parameter \( U \), which is allowed to vary with a
    uniform prior in the interval \( -4 \leqslant \log U \leqslant
    2 \),
    \item dust attenuation following the model of
    \citet{Salim2018}, parametrized by \( \delta \), the power-law
    index allowing for deviation from the \citet{Calzetti2000}
    attenuation curve, and \( B \), the \qty{2175}{\angstrom} bump
    strength, which are both allowed to vary with uniform priors
    in the intervals \( -1.6 \leqslant \delta \leqslant 0.4 \) and
    \( 0 \leqslant B \leqslant 10 \); furthermore, we allow the \(
    V \)-band attenuation, \( A_V \), to vary
    \qtyrange[range-open-phrase = {\text{from} }, range-phrase = {
    \text{to} }]{0.01}{8}{\mag}, and set the factor by which stars
    in birth clouds are more obscured, \( \eta \), to be in the
    range \numrange{1}{3},
    \item assume a lognormal star formation history (SFH)
    parametrized by the time at which star formation peaks, \(
    t_\text{max} \), and \( \sigma_\text{SFH} \), the FWHM of the
    SFH, allowing them to vary with uniform priors in the
    intervals \qtyrange{0.001}{15}{\giga\year} and
    \qtyrange{0.001}{20}{\giga\year}, respectively,
    \item allow the total stellar mass formed to be in the
    interval \( 1 \leqslant \log_{10}\left(\mast / \unit{\msun}
    \right) \leqslant 13 \), and the stellar and gas-phase
    metallicities to be in the range \( 0.001 \leqslant Z /
    \unit{\zsun} \leqslant 2.5 \).
\end{enumerate*}
To simplify the analysis, we adopt a \emph{best-case scenario} and
set the redshift of each subsample as the corresponding simulation
redshift. We list in \cref{tab:sedfittingparams} the free
parameters and priors used in the fitting process.

\subsubsection{Estimating physical parameters for the observational samples}
    \label{subsubsec:obssedfitting}

{{Based on the photometric catalogues from the NGDEEP and JADES
surveys (see \cref{subsec:ngdeep,subsec:jades}), we retrieve the
corresponding NIRCam fluxes and perform SED fitting of the
observational samples with the \texttt{BAGPIPES} code}}. The
fitting procedure mirrors that used for the simulated sample,
except for the model redshift. For the observational data, this
parameter is set as a Gaussian prior centred on the estimated
photometric redshift of each object, allowing deviations of up to
\( 3 \sigma \) in either direction, where \( \sigma = 0.25 \). As
noted above, these redshifts were derived using the \texttt{EAZY}
code by \citet{Leung2023} and \citet{Rieke2023} for the NGDEEP and
JADES samples, respectively. We refer the reader to those works
for details on their methodologies and any potential differences.
For both NGDEEP and JADES, we use photometry from the same six
NIRCam filters as in the simulated sample. Additionally, as the
JADES catalogue includes observations in the filters F090W, F335M
and F410M, we also consider them in this procedure when possible.

\section{Results} 
    \label{sec:results}

In this section, we present the results of the SED fitting for the
TNG50 sample and the NGDEEP and JADES observational samples. We
first show the performance of the SED fitting procedure in
recovering the intrinsic redshift and stellar mass of the
simulated galaxies. We evaluate this by computing commonly-used
performance metrics. Then, we present the \emph{UVJ} diagram for
the TNG50 sample and compare it with the same diagram for the
observational datasets.

\subsection{Recovering photometric redshifts from synthetic photometry}
    \label{subsec:recoveringz}

We quantify redshift recovery using three metrics: the mean bias
(\( \delta_z \)), the outlier fraction (\( \eta \)), and the
normalized median absolute deviation (NMAD; \(
\sigma_\text{\textsc{nmad}} \)). These are defined as follows:

\begin{enumerate}[leftmargin=*]
    \item The mean bias, \( \delta_z \), represents the average
    separation between the estimated and true values and is
    defined as:
    \begin{equation}
        \delta_z = \left\langle \frac{\Delta z}{1+z_\text{true}} \right\rangle,\label{eq:bias}
    \end{equation}
    where \( \Delta z = z_\text{phot} - z_\text{true} \) is the
    residual between the estimated photometric redshifts and the
    true values.
    
    \item The outlier fraction, \( \eta \), indicates
    the ratio of objects whose estimated redshifts deviate from
    the true values by a given threshold, and is defined as: 
    \begin{equation}
        \eta = \frac{N_\text{out}}{N},
    \end{equation}
    where \( N_\text{out} \) is the number of outliers, counted as
    those objects for which \( \left| \Delta z \right| /
    (1+z_\text{true}) > 0.15 \), and \( N \) is the sample size.
    
   \item The normalized median absolute deviation, \(
    \sigma_\text{\textsc{nmad}} \), denotes the dispersion of the
    residuals and is given by:
    \begin{equation}
        \sigma_\text{\textsc{nmad}} = 1.48 \times \text{median} \left( \left| \frac{\Delta z - \text{median}(\Delta z)}{1+z_\text{true}} \right| \right).
    \end{equation}
\end{enumerate}

\begin{table}
    \centering
    \caption{Performance metrics for the photometric redshift
    estimation of the TNG50 sample at \( z_\text{true} = 3, 4, 5\)
    and \( 6 \). The columns show the mean bias, the outlier
    fraction, and the normalized median absolute deviation (NMAD)
    for each bin. Redshift recovery is generally good when
    averaged over each subsample, but these metrics also indicate
    a slight underestimation of the true redshifts, with the
    greatest dispersion at \( z = 6 \).}
    \label{tab:photzmetrics}
    \begin{tblr}{
        width=\columnwidth,
        colspec={
        Q[si={table-format=1},l,co=1]
        Q[si={table-format=-1.4},c,co=1]
        Q[si={table-format=1.4},c,co=1]
        Q[si={table-format=1.3},r,co=1]
        },
        rowsep=0pt,
        row{1,2} = {guard},
        row{1} = {belowsep=0pt, abovesep=0pt}
    }
    \toprule
    Redshift & Mean bias & Outlier fraction & NMAD
    \\
    \midrule
    3 & -0.0014 & 0.0072 & 0.015 \\
    4 & -0.059 & 0.12 & 0.03 \\
    5 & -0.016 & 0.02 & 0.011 \\
    6 & -0.23 & 0.54 & 0.24 \\
    \bottomrule
    \end{tblr}
\end{table}

\begin{figure*}
    \includegraphics[width=\textwidth]{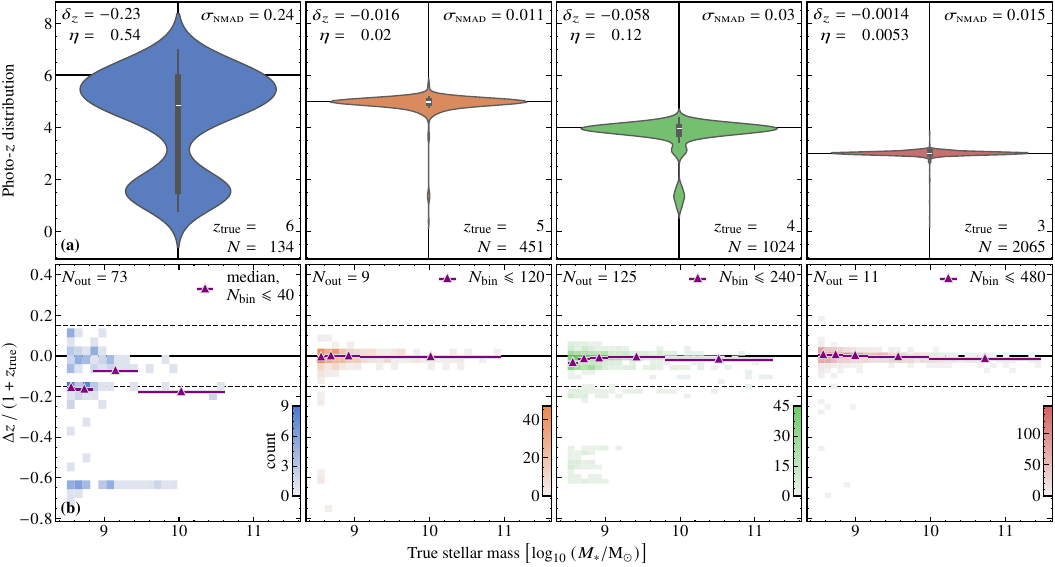}
    \caption{\textbf{(a)} Violin plots of the photometric redshift
    distributions of TNG50 galaxies at \( z_\text{true} = 6, 5,
    4\) and \( 3 \). Each panel shows mirrored kernel density
    estimates of the predicted redshifts. The horizontal line
    marks the true redshift; the inner box-and-whisker plot
    indicates with a thick bar the interquartile range (IQR; 25th
    to 75th percentile), with thin vertical lines extending to the
    most extreme point within \( 1.5 \times \text{IQR} \), and
    with a white line the median. Summary statistics are listed in
    each panel. Photometric redshifts are well recovered for \(
    z_\text{true} \leqslant 5 \) with near-zero bias and a low
    outlier rate. At \( z_\text{true} = 6 \), performance
    declines, with larger underestimations and more outliers.
    \textbf{(b)} Residuals of the estimated photometric redshifts
    versus the true stellar mass (\textsc{subfind} values). Purple
    triangles mark medians in bins of up to
    \numlist{40;120;240;480} objects for \(z_\text{true} = 6, 5,
    4\) and \( 3 \), respectively; horizontal bars show the bin
    widths. Dashed lines indicate the outlier threshold,
    \(\left|\Delta z\right| / (1+z_\text{true}) > 0.15\), with the
    number of outliers shown in each panel. Residuals remain
    largely flat with stellar mass, except at the low-mass end \(
    \left( \log_{10} \left( \mast / \unit{\msun} \right) \lesssim
    9\thinspace\text{\textendash}\thinspace10) \right) \), where
    they are the largest at all redshifts.}
    \label{fig:photzvstruez}
\end{figure*}

In \cref{fig:photzvstruez}, we present the estimated photometric
redshift distribution for each subsample, along with the residuals
as a function of the true stellar mass. We list performance
metrics for each snapshot in \cref{tab:photzmetrics}. In general,
as seen in the top row of \cref{fig:photzvstruez}, we find that
the estimated redshifts are well recovered, particularly for \( z
\leqslant 5 \) where the bias is around zero. {{At \( z = 4 \) and
\( 6 \), performance declines the most, with \( \eta = 0.12 \) and
\( 0.54 \), respectively}} indicating significant challenges in
recovering true redshifts {{at these snapshots}}. Additionally, at
each redshift, the estimations tend to underpredict (\( \delta_z <
0 \)) the true values, with the greatest dispersion of residuals
occurring at \( z = 6 \), where \( \sigma_\text{\textsc{nmad}} =
0.24 \). In the bottom row of \cref{fig:photzvstruez}, we show the
residuals as a function of the true stellar mass (\textsc{subfind}
value) for each snapshot. The purple triangles represent the
median value in bins that contain at most \numlist{480;240;120;40}
objects for the subsamples at \( z = 3, 4, 5\) and \( 6 \),
respectively, and the horizontal bars indicate the bin width. We
observe a mildly negative trend between the residuals and stellar
mass, particularly at the low-mass end \( \left( \log_{10}\left(
\mast / \unit{\msun} \right) \lesssim 9 \right) \), where the
scatter tends to be the largest, regardless of redshift. This
trend is most pronounced at \( z = 6 \), {{and to a less extent at
\( z = 4 \)}}, where redshifts are significantly underestimated
for a substantial fraction of objects. Additionally, the residual
distributions {{at these snapshots}} exhibit a {{multimodal}}
shape, with a {{prominent}} secondary peak at \( z_\text{phot}
\lesssim 2 \). This is in contrast to the other redshifts, where
residuals mostly show a flat trend with mass, as indicated by the
median values. {{We further discuss these findings in
\cref{sec:discussion}}}.

\subsection{Recovering stellar masses from synthetic photometry}
    \label{subsec:recoveringmasses}

We assess the accuracy of stellar mass recovery by calculating the
residual between the fitted and true values, defined as \( \Delta
\mast = \log_{10} \left( M_{\ast,\,\text{fit}} /
M_{\ast,\,\text{true}} \right) \). Here, \( M_{\ast,\,\text{fit}}
\) represents the estimated stellar mass, while \(
M_{\ast,\,\text{true}} \) denotes the corresponding true value. In
this case, the true stellar mass for each object is derived from a
projected stellar mass map created from the simulation particle
data. Each map matches the resolution of the images, and the true
value is computed within the same aperture used for the photometry
measurements. Additionally, we consider the following summary
statistics:

\begin{enumerate}[leftmargin=*]
    \item The population mean, \( \mu \), which provides an
    overall measure of the mass recovery accuracy across each
    subsample.
    \item The standard deviation, \( \sigma \), which quantifies
    the dispersion of the residuals and indicates the level of
    scatter in the distribution.
    \item The mean value of the mass-binned median residuals, \(
    \bar{\mu} \), which gives an estimate of the systematic bias
    in the mass recovery. This is computed by taking the median of
    the residuals in bins of width \( \delta \log_{10} \left(
    \mast / \unit{\msun} \right) \approx \num{0.8} \) and
    averaging across all bins.
    \item The half-mean of the mass-binned \qty{68}{\percent}
    range of the residuals, \( \bar{\sigma} \), which provides an
    estimate of the scatter in the mass recovery. This is computed
    by taking the \qty{68}{\percent} range (i.e. the difference
    between the \num{84}th and \num{16}th percentiles) of the
    residuals in each bin, averaging across all bins, and dividing
    the result by two. We employ the same bin width as for \(
    \bar{\mu} \).
\end{enumerate}

\begin{figure}
    \includegraphics{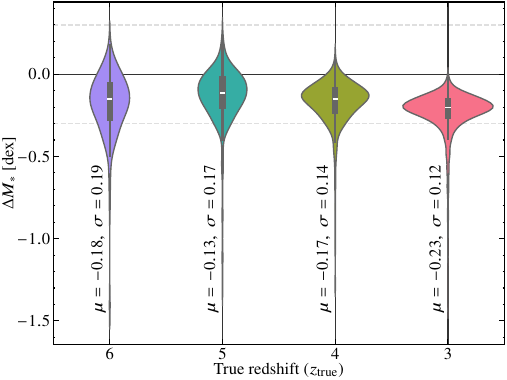}
    \caption{ Distribution of stellar mass residuals, \( \Delta
    \mast \), for the TNG50 sample at \( z_\text{true} = 6, 5, 4
    \) and \( 3 \). The violin plots show the kernel density
    estimates of the residuals, while the box-and-whisker plots
    inside represent the interquartile range (IQR), with the white
    line indicating the median. The horizontal solid line marks
    zero residual, and the dashed lines indicate \qty{\pm
    0.3}{dex} (about a factor of 2). Stellar masses are mostly
    underestimated, but are generally recovered within
    \qty{0.3}{dex}. The population mean, \( \mu \), becomes less
    negative and the distributions show increased scatter at
    higher redshifts.}
    \label{fig:massrecovery1d}
\end{figure}

In \cref{fig:massrecovery1d}, we plot the distribution of \(
\Delta \mast\) for each subsample, along with summary statistics.
In general, when considering the full sample, the stellar masses
are recovered reasonably well, with distribution peaks above
\qty{-0.3}{dex} (a factor of 2) and a population mean of at most
\qty{-0.23}{dex}. The scatter in the residuals is also low, with a
standard deviation of at most \qty{0.19}{dex}. As redshift
decreases, the residual distribution becomes narrower and the
underestimation of the masses increases, with the largest \( \mu
\) at \( z_\text{true} = 3 \).

\begin{figure*}
    \includegraphics[]{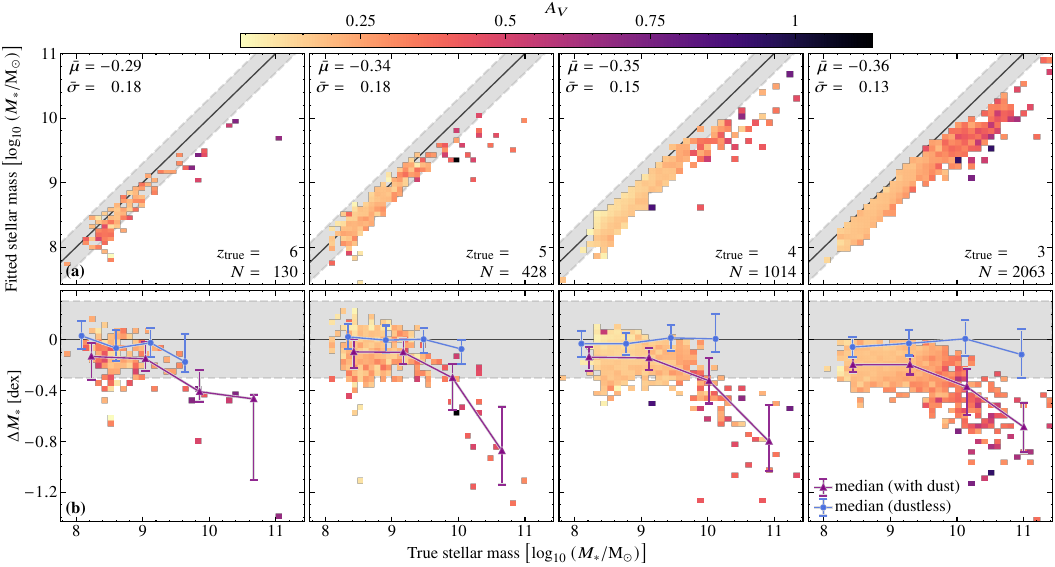}
    \caption{\textbf{(a)} Fitted versus true stellar masses for
    the TNG50 subsamples at \( z_\text{true} = 6, 5, 4 \) and \( 3
    \). The colour scale shows the predicted value of
    \emph{V}-band attenuation for objects in each bin. The solid
    line indicates the one-to-one relation, while the dashed lines
    mark offsets by a factor of 2. \textbf{(b)} Mass residuals as
    a function of true stellar mass. Purple triangles mark median
    values, with error bars spanning the 16th to 84th percentiles;
    horizontal dashed lines indicate \qty{\pm0.3}{dex}. For
    low-mass galaxies  \( \left( \log_{10}\left( \mast /
    \unit{\msun} \right) \lesssim 9 \right) \), residuals are flat
    with underestimations up to a factor of \num{2}. At higher
    masses \( \left( \log_{10}\left( \mast / \unit{\msun} \right)
    \gtrsim 10 \right) \), offsets become increasingly negative
    even exceeding \qty{-1}{dex}. These high-mass galaxies with
    the largest offsets, exhibit the highest predicted
    \emph{V}-band attenuation values. The mass-binned mean \(
    \bar{\mu} \) peaks at \( z = 3 \), while the scatter \(
    \bar{\sigma} \) decreases with lower redshift. In contrast,
    using dustless synthetic photometry (blue line) nearly
    eliminates the systematics, with stellar masses recovered
    within \qty{0.2}{dex} or less, {{indicating that dust
    attenuation is the main driver of the discrepancies. This
    suggests that the strength of dust attenuation is
    underestimated in the SED fitting, especially at high
    masses.}}}
    \label{fig:massrecovery2d} 
\end{figure*}

In \cref{fig:massrecovery2d}, we present the estimated stellar
mass and the mass residuals as a function of the true stellar mass
(derived from a projected stellar mass map) for each subsample. At
the low-mass end \( \left( \log_{10} \left( \mast / \unit{\msun}
\right) \leqslant 9.0 \right) \), we observe that stellar masses
are generally underestimated within a factor of \num{2},
regardless of redshift, and the residuals remain relatively flat
with mass. For masses \( \log_{10} \left( \mast / \unit{\msun}
\right) \geqslant 10 \), the offset between the estimated and true
values becomes more significant, especially at \( z_\text{true} =
3 \), where  \( \bar{\mu} \) is \qty{-0.37}{dex} and a substantial
fraction of objects have masses underestimated by more than
\qty{0.4}{dex} (a factor of \num{2.5}). This trend is slightly
less pronounced at higher redshifts, where \( \bar{\mu} \) is
smaller. From this figure we also find that, in general, high-mass
galaxies exhibit the largest values of predicted \emph{V}-band
attenuation. 

{{\cref{fig:massrecovery2d} also shows the result of estimating
stellar mass using synthetic observations excluding the effects of
both a dust distribution (no radiative transfer) and emission
lines from star-forming regions (no \texttt{MAPPINGS-III}), and
without employing a dust attenuation model or emission lines in
the SED fitting}}. These results are shown as blue lines in the
same figure. Overall, we find better recovery of stellar masses,
with a flatter residual distribution and offsets of at most
\qty{0.2}{dex}, regardless of true stellar mass. A likely
explanation for the discrepancies between these two cases is
geometric in nature: high-mass galaxies, being typically more
concentrated, exhibit larger dust column densities and are
therefore more strongly affected by dust attenuation
{{\citep[e.g.][]{Hamed2023}}}. {{In addition, high-mass galaxies
have larger dust content per unit stellar mass \citep[e.g.][find
that both the UV attenuation and infrared excess increase with
stellar mass]{Pannella2015,McLure2018,Shivaei2020}, which could
further contribute to the trend between mass residuals and stellar
mass when attenuation is included.}} This is consistent with the
estimated \( A_V \) values shown in the figure.

\begin{figure*}
    \includegraphics[]{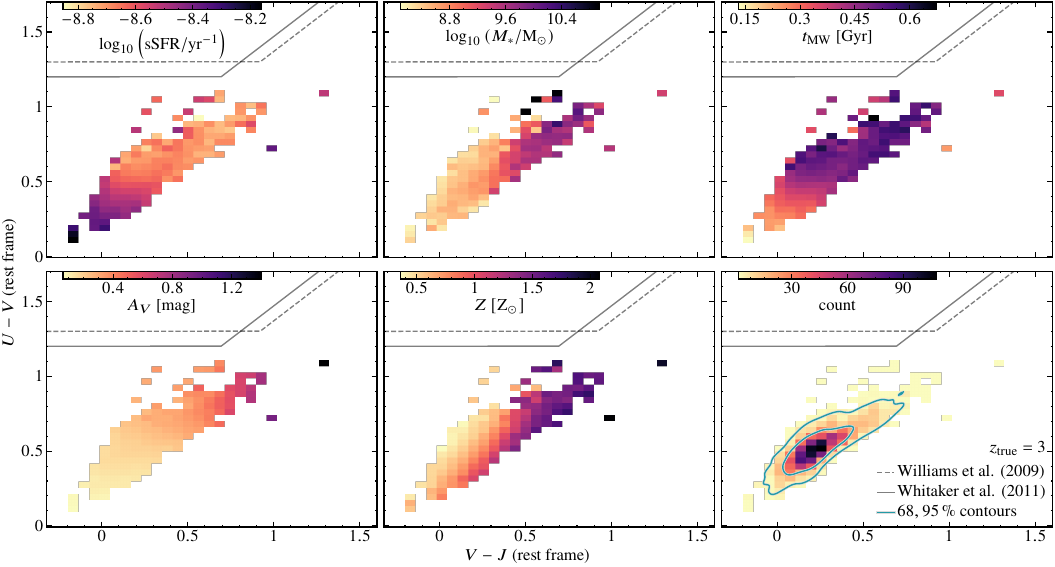}
    \caption{The \emph{UVJ} diagram for the TNG50 sample at \( z =
    3 \). The plot shows best-fit predicted \( U - V \) and \( V -
    J \) for the sample, colour-coded (from left to right) by the
    specific star formation rate (sSFR), stellar mass,
    mass-weighted age \( (t_\text{MW}) \), \emph{V}-band
    attenuation, and metallicity. The dashed and solid lines
    represent the \citet{Williams2009} and \citet{Whitaker2011}
    boundaries between quiescent and star-forming galaxies,
    respectively. According to these criteria, no objects fall
    into the quiescent region. Clear trends are seen: \( U - V \)
    colours are inversely related to sSFR (redder colours
    correspond to lower sSFRs); both stellar mass and
    mass-weighted age correlate with  \( {{V - J}} \) colour, with
    more massive and older galaxies being redder; and
    \emph{V}-band attenuation increases with redder \emph{UVJ},
    with the bluest objects having the lowest \( A_V \) values,
    while those more attenuated are located closer to the
    top-right corner. These trends are consistent with previous
    studies at lower redshifts. The final panel shows the
    distribution of objects in this plane, with green contours
    indicating the \num{68} and \qty{95}{\percentt} levels.
    {{Colours are expressed in the AB system.}}}
    \label{fig:uvjdiagramtng50z=3}
\end{figure*}

\subsection{The \emph{UVJ} diagram of TNG50 high-redshift
galaxies}
    \label{subsec:uvjdiagramtng50}

In \cref{fig:uvjdiagramtng50z=3}, we present the \emph{UVJ}
diagram for the TNG50 sample at \( z = 3 \). The diagram shows the
predicted \( U-V \) and \( V-J \) colours of each object,
colour-coded by the {{SED-inferred}} specific star formation rate
(sSFR), stellar mass, mass-weighted age \( (t_\text{MW}) \),
metallicity, and \emph{V}-band attenuation. The dashed line
represents the \citet{Williams2009} division between quiescent and
star-forming galaxies, while the solid line indicates the
\citet{Whitaker2011} boundary. Notably, these selection boundaries
result in practically no objects classified as quiescent in the
TNG50 sample, although they were originally proposed up to \(
z\sim2.5 \). We note that \( U - V \) and \( V - J \) colours
anticorrrelate with sSFR, in the sense that objects with larger \(
U - V \) have lower sSFRs, approaching the quiescent region.
Moreover, stellar mass and, to some extent, the mass-weighted age
and metallicity show a correlation with the \( {{V - J}} \)
colour, where the most massive and oldest galaxies tend to display
redder colours. Lastly, the {{SED-inferred}} \emph{V}-band
attenuation also correlates with \( U - V \) and \( V - J \): as
expected, the bluest objects show the lowest values of \( A_V \),
while those with the largest \( A_V \) values locate closer to the
top-right corner and almost parallel to the diagonal boundaries.
These trends are consistent with previous studies
\citep[e.g.][]{Belli2017,Schreiber2018,Leja2019,Baes2024a}.

\begin{figure*}
    \includegraphics{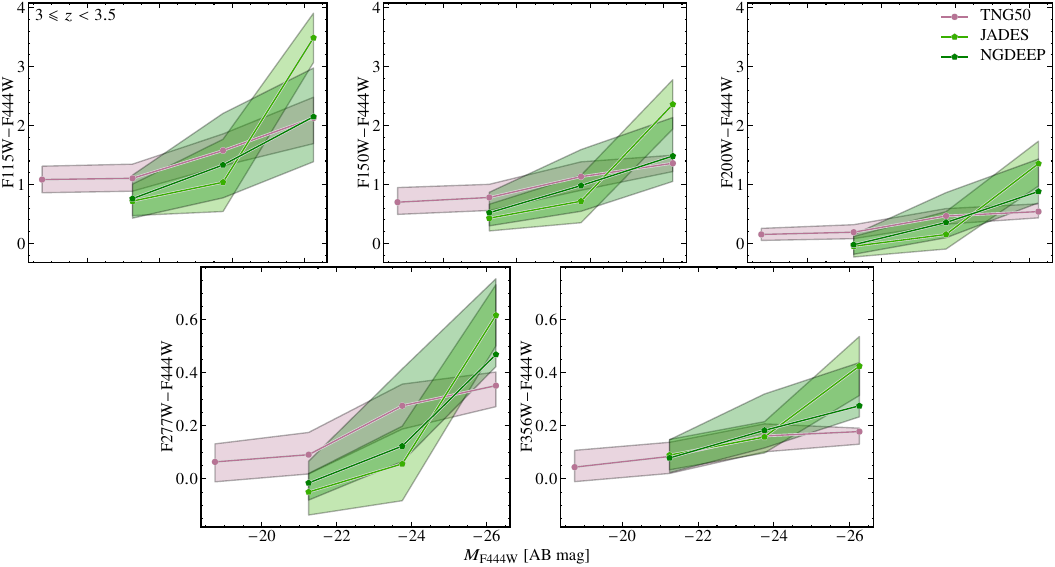}
    \caption{Median colour indices against absolute magnitude for
    the TNG50 sample at \( 3 \leqslant z < 3.5 \) compared to
    NGDEEP and JADES observations. The vertical axis represents
    colours with respect to the F444W filter, while the horizontal
    axis is the absolute magnitude in F444W. The shaded regions
    indicate the range from the 16th to the 84th percentiles of
    the distributions. We find that TNG50 broadly aligns with the
    trends observed in NGDEEP and JADES, with median colours
    falling within similar regions. However, at fixed absolute
    magnitudes, the TNG50 sample generally has redder colours
    compared to observations, except for the brightest galaxies in
    the sample. {{Colours are expressed in the AB system.}}}
    \label{fig:colormag}
\end{figure*}

\begin{figure*}
    \includegraphics{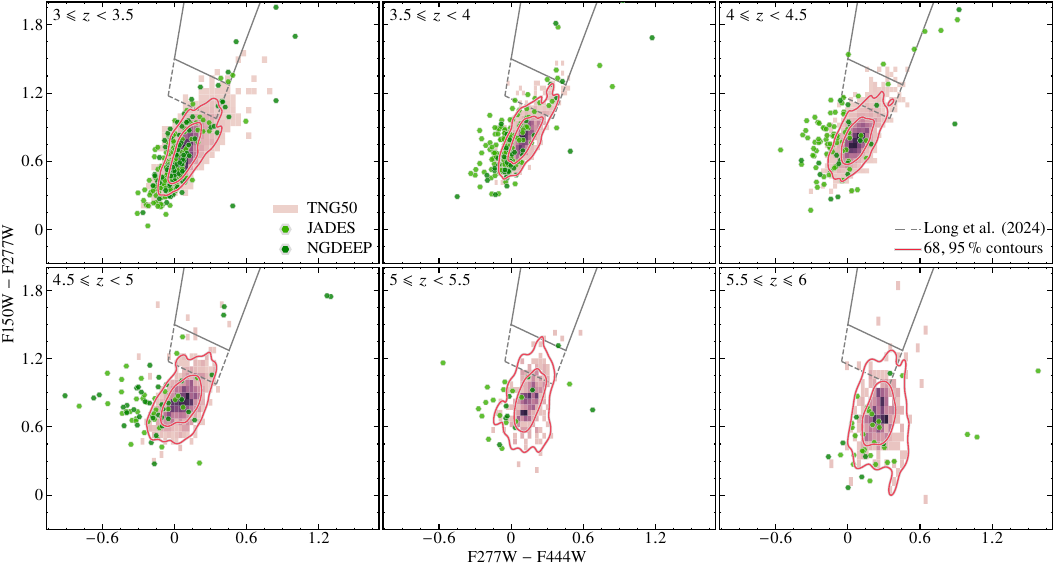}
    \caption{The \( \text{F150W} - \text{F277W} \) against \(
    \text{F277W} - \text{F444W} \) colour-colour diagram for the
    TNG50 sample at different redshift bins between \( z = 3 \)
    and \( 6 \) compared to NGDEEP and JADES observations. The
    solid (dashed) gray line represents the strict (flexible)
    boundary between quiescent and star-forming galaxies proposed
    by \citet[]{Long2024}. The red contours indicate the \num{68}
    and \qty{95}{\percentt} of the simulated distribution. In this
    colour space we find broad agreement between the simulated and
    observed samples, but with a tendency for the TNG50 objects to
    exhibit redder colours than observations regardless of
    redshift. Additionally, both samples contain a small fraction
    of objects denoted as quiescent according to the extended
    \citet[]{Long2024} criteria. It is important to note that this
    extended region is susceptible to contamination from
    post-starburst and star-forming galaxies, with the latter
    constituting the dominant population in both the simulated and
    observed samples. {{Colours are expressed in the AB system.}}}
    \label{fig:long24diagram}
\end{figure*}

\begin{figure*}
    \includegraphics{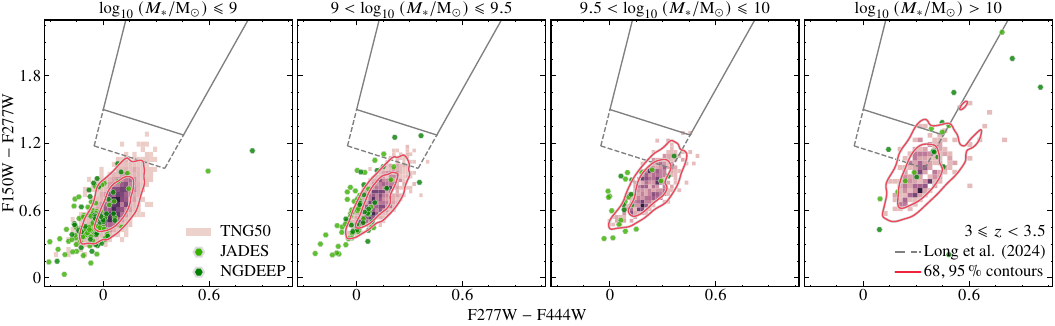}
    \caption{The \( \text{F150W} - \text{F277W} \) against \(
    \text{F277W} - \text{F444W} \) colour-colour diagram for
    different mass bins at \( 3 \leqslant z < 3.5 \) for the TNG50
    sample compared to NGDEEP and JADES observations. The
    simulated and observed samples show a similar distribution of
    colours, with the TNG50 sample generally exhibiting redder
    colours in both axes, particularly at intermediate and low
    masses. For both samples, the most massive galaxies tend to
    occupy in larger proportions the extended quiescent region
    defined by \citet{Long2024}. {{Colours are expressed in the AB
    system.}}}
    \label{fig:long24_tng50_vs_jwst_by_mass}
\end{figure*}

\begin{figure*}
    \includegraphics{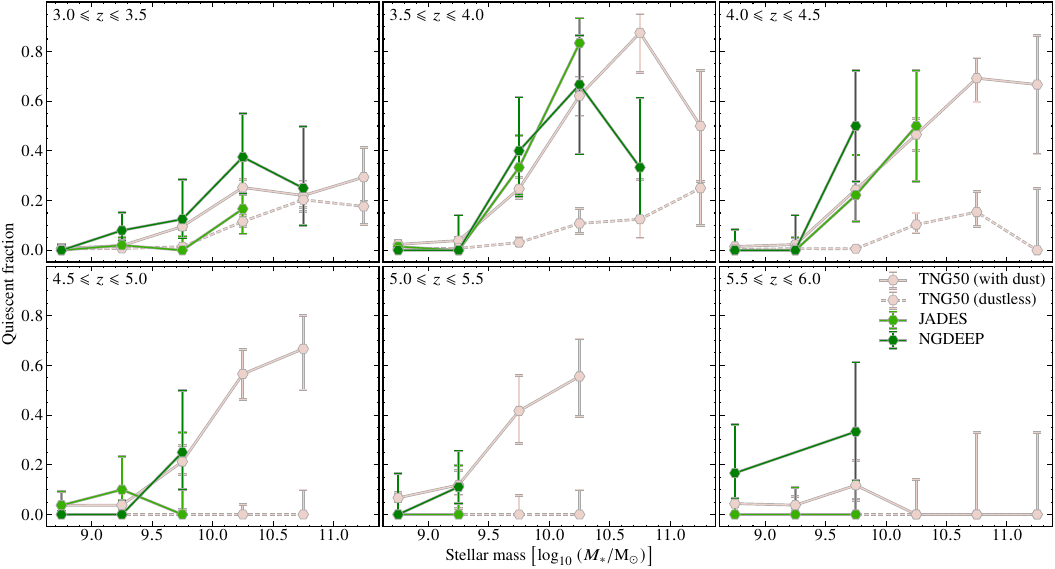}
    \caption{The mass-dependent quiescent fraction for TNG50,
    JADES, and NGDEEP at \( 3 \leqslant z \leqslant 6 \).
    Quiescent candidates were selected using the colour-based
    criterion from \citet[]{Long2024}. The simulation estimate is
    broadly consistent with observations within statistical
    uncertainty up to \( z \sim 5 \). For comparison, the dustless
    TNG50 fraction is also shown, following a strict monotonic
    trend for the lower redshift bins. Error bars indicate
    binomial proportion confidence intervals.}
    \label{fig:quies_frac}
\end{figure*}

\subsection{Comparing the colours of high-redshift TNG50 galaxies with JWST observations}
    \label{subsec:colorcomparison}

We compare the synthetic TNG50 images to observations by first
plotting median colour -- for various NIRCam filters -- against \(
\text{F444W} \) absolute magnitude. Secondly, we compare
colour-colour diagrams for NIRCam filters and the best-fit
\emph{UVJ} diagrams in the rest frame of the TNG50 data with those
of NGDEEP and JADES.

In \cref{fig:colormag}, we show median colour indices versus
absolute magnitude for the TNG50 sample alongside the NGDEEP and
JADES observations at \( 3 \leqslant z < 3.5 \). The vertical axis
represents the median colour between any of the F115W, F150W,
F200W, F277W, and F356W filters and the F444W filter, while the
horizontal axis shows the absolute magnitude in the F444W filter,
and the shaded regions indicate the extent from the 16th to the
84th percentiles of the distributions. We note that these plots
are based on intrinsic measured quantities. As expected, the
NGDEEP and JADES colour trends are similar with each other, both
becoming redder as the absolute magnitude decreases and generally
overlapping within their respective 16th to 84th percentile
ranges, however, NGDEEP galaxies tend to have slightly redder
colours than JADES. At the same time, the TNG50 sample is broadly
consistent with observations, with median colours falling within
similar regions. Nonetheless, at fixed absolute magnitudes, the
TNG50 sample shows a tendency toward redder colours relative to
observations, except for the brightest galaxies, where the
trend reverts.

In \cref{fig:long24diagram,fig:long24_tng50_vs_jwst_by_mass}, we
present the colour-colour selection diagram introduced by
\citet{Long2024} for identifying quiescent galaxies at \( 3 < z <
6 \), first across different redshift bins and then subdivided by
stellar mass at \( 3 \leqslant z < 3.5 \). This diagram uses the
\( \text{F150W} - \text{F277W} \) versus \( \text{F277W} -
\text{F444W} \) colours to distinguish quiescent from star-forming
galaxies. We find that both the TNG50 and JWST samples broadly
follow similar distributions in this colour space, with the
simulated galaxies generally exhibiting slightly redder colours as
in the colour-absolute magnitude diagrams. Using the stringent
selection cut from \citet[eq. (1) there]{Long2024}, we identify a
few quiescent candidates in both samples, which increases if the
flexible selection cut (eq. (2) there) is applied. As has been
pointed out in such work, the latter choice could increase the
degree of contamination from star-forming galaxies but also
improve the detection of post-starbursts, which can be missed from
traditional selection schemes.

Using the flexible colour cut from \citet[]{Long2024}, we select
quiescent candidates in all samples and calculate the quiescent
fraction as a function of stellar mass across redshift. In
\cref{fig:quies_frac}, we show the mass dependence of this
quantity at different redshift bins in the \( 3 \leqslant z
\leqslant 6\) interval for TNG50, JADES, and NGDEEP. At \( 3
\leqslant z \lesssim 5\), the simulation estimate is broadly
consistent with observations within statistical uncertainty, in
contrast with \citet[]{Weller2025}, who did not include the
effects of dust attenuation in their analysis. For comparison, we
also apply the same selection to TNG50 dustless synthetic
photometry (dashed lines), finding that the quiescent fraction
rises monotonically with stellar mass up to \qty{\sim
1e11}{\msun}, whereas the dust-aware estimate shows a
non-monotonic trend. Similar behaviour is seen up to \( z \sim 5
\), with observational and TNG50 dust-aware fractions increasing
up to \( \log_{10}\left(\mast / \unit{\msun}\right) \sim
\numrange[range-phrase=\text{~--~}]{10.25}{10.75} \) and attaining
maximum values between sixty and eighty percent, before flattening
or declining. At \( z \gtrsim 5 \), the observational samples
become insufficient, impeding to draw robust conclusions.

\begin{figure*}
    \includegraphics{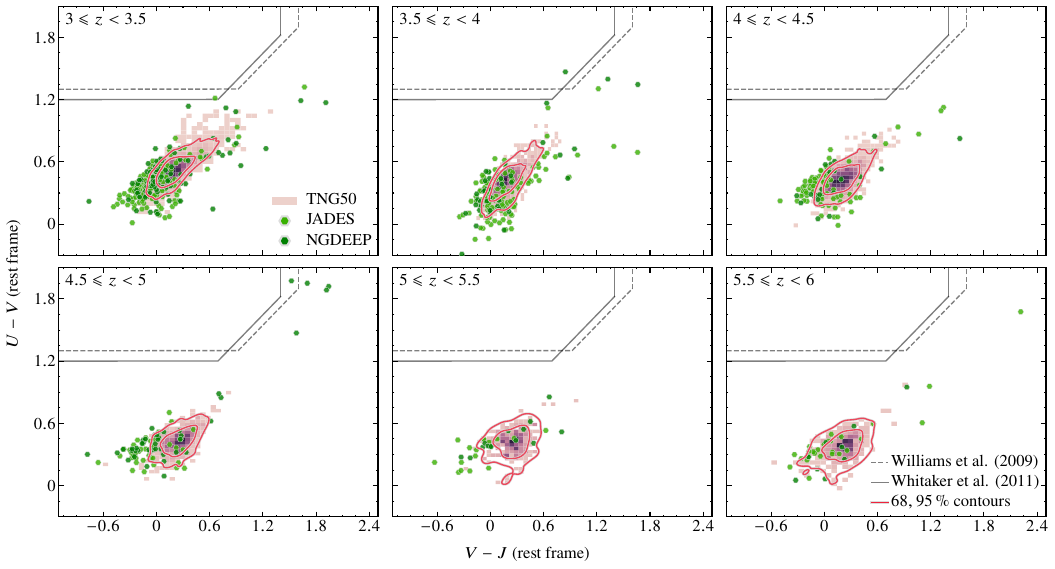}
    \caption{Comparison of the best-fit \emph{UVJ} diagrams for
    the TNG50, NGDEEP, and JADES samples at different redshift
    bins between \( z = 3 \) and \( 6 \). We find that the
    simulated sample is broadly consistent with the observations
    in both \( U - V \) and \( V - J \) colours. As redshift
    decreases, the distributions become more spread out in the red
    direction of the diagram, with objects mainly approaching the
    region of dusty star-forming galaxies in the top-right corner.
    Regardless of redshift, the main discrepancy between the
    simulation and observations is that the TNG50 sample peaks at
    redder in \( V - J \) colour. {{Colours are expressed in the
    AB system.}}} 
    \label{fig:uvjcomparison}
\end{figure*}

\begin{figure*}
    \includegraphics{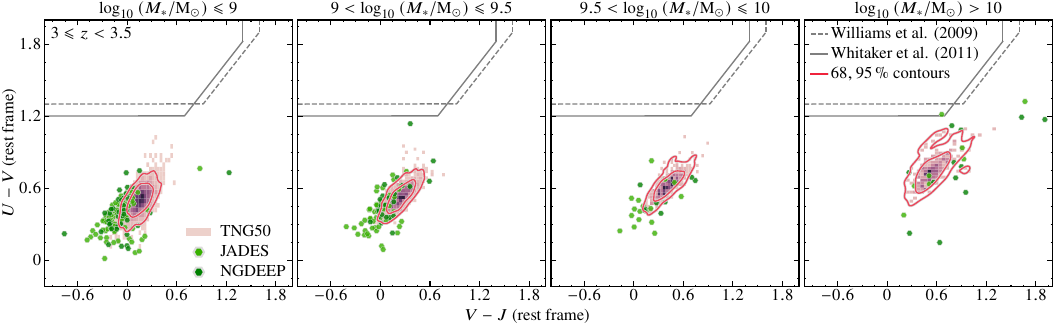}
    \caption{Mass-dependent comparison of the best-fit \emph{UVJ}
    diagram for the TNG50, NGDEEP, and JADES samples at \( 3.5
    \leqslant z < 3 \). The synthetic distributions are compatible
    with the expected colours of observations, as they exhibit a
    similar range of colours and become more red in both \( U - V
    \) and \( V - J \) as mass increases. In general, the TNG50
    sample shows redder colours in \( V - J \), with the
    discrepancy being more pronounced in both colours
    (observations are bluer) at intermediate masses. At the
    high-mass end, the simulation data do not significantly extend
    into the redder region, and there is no sufficient evidence to
    suggest that there is a significant population of
    dust-reddened galaxies in the observations. {{Colours are
    expressed in the AB system.}}}
    \label{fig:uvj_tng50_vs_jwst_by_mass}
\end{figure*}

Complementary to the previous, we show in
\cref{fig:uvjcomparison}
the best-fit \emph{UVJ} diagram for all samples at different
redshift bins between \( z = 3 \) and \( 6 \). Both observational
datasets predominantly consist of objects below the
\citet{Williams2009} and \citet{Whitaker2011} boundaries, with
some objects extending into the so-called dusty star-forming
region in the top-right corner, and little to no galaxies crossing
into the quiescent region. Compared to NGDEEP, JADES galaxies are
more concentrated in the star-forming region, a trend that is also
observed in the TNG50 sample. However, as in the colour-absolute
magnitude diagrams, the TNG50 distributions peak at a slightly
redder \( V - J \) colour, regardless of redshift. At \( 3
\leqslant z < 3.5 \), this is consistent with the behaviour shown
for the intrinsic \( \text{F200W} - \text{F444W} \) index in
\cref{fig:colormag}, as at \( z \sim 3 \) this colour has the most
overlap with the \( V \) and \( J \) filters. In
\cref{fig:uvj_tng50_vs_jwst_by_mass}, we divide the \( 3 \leqslant
z < 3.5\) range into four mass bins and compare the respective
colours. Overall, we find that the best-fit \emph{UVJ} colours of
the TNG50 sample are compatible with the expected colours of
high-redshift galaxies, as the synthetic distributions exhibit a
similar range of colours to the observations, and also become more
red in both \( U - V \) and \( V - J \) colours as mass increases,
moving closer to the right-top corner. However, some differences
appear in the synthetic distributions regardless of mass. In
particular, the TNG50 sample peaks at redder colours in both
directions at intermediate and low masses, and to a less extent it
is slightly bluer than observations at the high-mass end.

\section{Discussion}
    \label{sec:discussion}

In \cref{subsec:recoveringz}, we demonstrated that photometric
redshifts in the TNG50 sample are accurately recovered, especially
at \( z \leqslant 5 \), where \qty{>90}{\percentt} of the
estimations fall within \num{\pm0.2}. {{However, at \( z = 6 \),
and to a lesser extent at \( z = 4 \), there is a noticeable
underestimation of redshifts, with a significant fraction of
outliers, and multimodal photo-\( z \) distributions rather than
random scatter around the true values. This suggests that certain
combinations of true redshifts and broadband filters can lead to
multiple solutions that fit the photometry equally well. Such
degeneracies arise from the coarse sampling of the SED by only six
NIRCam bands, in tandem with the discrete template set in
\texttt{EAZY}. Upon examination of problematic cases, we confirm
that the fits worsen in the bluer regions of the SEDs and the
redshift probability distributions typically have at least two
distinct peaks, with a dominant low-redshift solution and a
secondary one near the true redshift. This pattern suggests a
confusion between spectral features like the \qty{4000}{\angstrom}
break (falling in F115W at \( z \sim 2 \) and in F277W at \( z
\sim 6 \)) or degeneracies between the Lyman and
\qty{4000}{\angstrom} breaks, where a red, low-\( z \) object
exhibiting a \qty{4000}{\angstrom} break can mimic a young,
high-\( z \) system with depleted blue flux due to Lyman-\( \alpha
\) absorption). Since we only probe the red side of the Lyman
break at the redshifts considered here, the fitting code might
confuse these features, leading to incorrect redshift estimates.
While the relatively high outlier fractions at \( z = 4 \) and \(
6 \) can be a cause for concern in the absence of spectroscopic
data for e.g. estimating galaxy abundances, it is important to
note that:
\begin{enumerate*}[leftmargin=*, label=(\emph{\roman*})]
    \item our global statistics are comparable to those reported
    in recent observational campaigns. For example,
    \citet{Merlin2024} report an outlier fraction of
    \qty{\sim6}{\percentt} for an aggregate of JWST surveys,
    whereas \citet{Shuntov2025} find a lower value of
    \qty{\sim2}{\percentt} for COSMOS-Web \citep[][]{Casey2023},
    and
    \item our analysis relies on six NIRCam bands for the fitting
code. Including additional data from bluer filters than F115W
would likely alleviate these issues by better constraining the
blue side of the SED and thus reducing the number of degenerate
solutions. This is consistent with the findings of
\citet{Bisigello2017,Kauffmann2020}, who performed similar
analyses to ours using mock photometry, and showed that such
additions can drastically reduce outlier fractions over a wide
redshift range. 
\end{enumerate*}
}} An alternative possibility is that the spectral templates are
suboptimal at \( z \gtrsim 6 \), since the observed colours of
real galaxies may differ from those in the models
\citep[e.g.][]{Finkelstein2022}. In this regard,
\citet{Larson2023} and \citet{Hainline2024} have proposed the use
of new template sets that better match the colours of
high-redshift galaxies, and that show improved recovery. Testing
this on the TNG50 sample at \( z = 6 \) using the recommended set
for \( z \sim 4 - 7 \) from \citet{Larson2023} yielded only a
slight reduction in the number of low-redshift solutions, with no
significant improvement in overall performance (see
\cref{subsec:appendix_photz}). This suggests that TNG50 galaxies
at \( z = 6 \) do not suffer from significant template mismatches,
and instead the noisier flux from the bluer filters {{and the
limited wavelength coverage are the main limitations}}.

Regarding the stellar mass recovery, we showed in
\cref{subsec:recoveringmasses} that SED fitting with six-point
NIRCam photometry recovers \qty{>95}{\percent} of stellar masses
within \qty{0.5}{dex} for the TNG50 sample at \( 3 \leqslant z
\leqslant 6 \) (see \cref{fig:massrecovery1d}). Despite this,
stellar masses are systematically underestimated regardless of
redshift. At stellar masses \( \log_{10} \left(\mast /
\unit{\msun}\right) \lesssim 10 \), the offset is mostly flat with
a median of about \qty{-0.2}{dex}, while at higher masses
residuals become more negative, with some underestimations greater
than an order of magnitude. As the inferred physical properties of
galaxies might be sensitive to the assumptions made during the
fitting procedure \citep[see e.g.][]{Conroy2013}, we consider
several plausible causes.

At higher masses, more negative offsets correlate with higher
inferred \emph{V}-band attenuation (see
\cref{fig:massrecovery2d}), suggesting that compact {{or in
general dustier galaxies}} are more likely to be misidentified as
lower mass systems. A second factor could be a \emph{nebular
bias}, where the code overestimates the contribution of emission
lines from younger stars, inflating equivalent widths and
underestimating the mass of the galaxy, as the contribution of
older stars is not properly included. \citet{Cochrane2025}
observed this effect using synthetic spectra at \( 5 \leqslant z
\leqslant 10 \) from the SPHINX simulations, and underestimating
masses by up to \qty{0.5}{dex} at \( z =
\numrange[range-phrase=\text{~--~}]{5}{6}\) and \( 9 \lesssim
\log_{10} \left(\mast / \unit{\msun}\right) \lesssim 10 \),
similar to our findings. A third factor could be
\emph{outshining}, where younger stars dominate the continuum thus
masking older populations, and leading to lower mass estimates.
\citet{Narayanan2024} demonstrated that this effect dominates at
\( z \gtrsim 7 \), largely independent of the assumed star
formation history. {{Given that, when excluding dust and nebular
emission from the synthetic photometry, we find nearly unbiased
mass estimates regardless of true stellar mass or redshift (blue
lines in \cref{fig:massrecovery2d}), this strongly suggests that
dust attenuation is the main driver of the mass underestimation,
rather than outshining, although the effect of nebular emission
cannot be completely ruled out as our fiducial images contain both
components. This is clearer at higher masses: as mass increases,
galaxies tend to be dustier but the fitting underestimates the
attenuation, inferring lower intrinsic fluxes, and thus larger
mass underestimations.}}

Lastly, the choice of priors always influence parameter
estimation. We made standard assumptions about them, adopting a
flexible dust model and a parametric SFH. Although parametric SFHs
can underperform at lower redshifts \citep[e.g.][]{Lower2020} and
non-parametric SFHs can alleviate tensions between theory and
observations \citep[e.g.][in the context of the star-forming main
sequence]{Nelson2021}, we find this is not the case here. The
level of uncertainty we report is consistent with studies using
more complex SFHs in similar redshift and mass ranges
\citep[e.g.][]{Cochrane2025}, suggesting that non-parametric SFHs
are not necessary in this regime,  as SFHs are expected to be
mostly rising. In any case, repeating the fits with different dust
and SFH priors did not significantly change the trends we found
(see \cref{subsec:appendix_masses}).

In \cref{subsec:uvjdiagramtng50}, we presented expectations for
the \emph{UVJ} diagram in TNG50. \cref{fig:uvjdiagramtng50z=3},
shows the sample at \(z = 3\) coloured by several SED-inferred
properties. First, according to commonly used criteria -- tuned up
to \( z \sim 2.5 \) -- there are no objects in the passive region
of the diagram, with the sample being dominated by star-forming
galaxies. This is expected as galaxies typically transition into
the passive region at \( z \lesssim 2.5 \), and quiescent
candidates remain low at \( z \gtrsim 3 \)
\cite[e.g.][]{Valentino2023}. Secondly, we find sensible
correlations between \emph{UVJ} and galaxy properties, such as the
anti-correlation between colours and sSFR, and the correlation
between stellar mass and, to a lesser degree, the mass-weighted
age with \(U - V\) colour. Additionally, we observe a strong
correlation between \emph{V}-band attenuation and \emph{UVJ}, with
the bluest objects having the lowest \(A_V\) values, and also find
that redder objects tend to have the largest metallicities. This
is in agreement to similar studies at lower redshifts.
\citet{Baes2024a} directly estimated the \emph{UVJ} diagram from
the TNG50 simulation at \(z = 0\), reporting trends consistent
with ours.

In
\cref{fig:colormag,fig:long24diagram,fig:long24_tng50_vs_jwst_by_mass},
we compared observer-frame photometric measurements between TNG50
and JWST data, by first plotting colour indices against absolute
magnitude and then considering a recent colour-colour diagram
proposed by \citet[]{Long2024}. Overall, we found that the
synthetic sample broadly follows the observations, showing similar
median trends in the colour-magnitude space (\cref{fig:colormag})
and occupying comparable loci in the \( \text{F150W} -
\text{F277W} \) versus \( \text{F277W} - \text{F444W} \) diagram
(\cref{fig:long24diagram,fig:long24_tng50_vs_jwst_by_mass}). The
main difference, however, is that the TNG50 sample is slightly
offset, generally showing redder colours than observations for the
majority of the population and NIRCam filters. This colour
discrepancy is observed across the entire redshift range
considered in this work. As these plots are based on direct fluxes
from radiative transfer post-processing, it is plausible
discrepancies arise from this procedure. For the simulated sample,
the total dust mass in a galaxy is set by the metal mass scaled by
a constant dust-to-metals ratio, which is a simplification as full
dust physics is not included in TNG50. Indeed, \citet{Akins2022}
studied the \emph{UVJ} diagram of galaxies at \( z = 2 \) in the
SIMBA cosmological simulation, finding that, compared to an
explicit dust model, assuming a constant dust-to-metals ratio
leads to redder colours in both directions of the \emph{UVJ}
diagram, and thus fails to fully match the observed colours of
galaxies. The offset we see in the colour-magnitude diagram could
be attributed to the same effect, since the synthetic \(
\text{F200W} - \text{F444W} \) colour is the one that shows the
most overlap with the \( V \) and \( J \) filters.

In \cref{fig:quies_frac}, we showed the mass-dependent quiescent
fraction at \( 3 \leqslant z \leqslant 6 \) for TNG50, JADES, and
NGDEEP, finding that the simulation estimate aligns, within
uncertainty, with observational trends up to \( z \sim 5 \).
Notably, for the dustless images we find that this quantity
increases monotonically with mass up to \qty{\sim 1e11}{\msun} and
peaks around twenty per cent, but for the dust-aware sample, the
trend is non-monotonic, attaining greater values and flattening or
declining at the high-mass end. This non-monotonic behaviour at \(
z \lesssim 5 \) is also seen in JADES and NGDEEP, which show
similar trends in this mass range. We attribute this to two main
channels for the reddening of galaxies:
\begin{enumerate*}[leftmargin=*, label=\emph{(\roman*)}]
    \item stellar evolution, which results in more massive
    galaxies having older stellar populations and is expected to
    increase monotonically with mass, and
    \item dust attenuation, which is expected to significantly
    increase around \qty{\sim 1e10}{\msun} and decline at the
    high-mass end due to the lack of cold gas in more massive
    systems.
\end{enumerate*}

In \cref{fig:uvjcomparison}, we presented the \emph{UVJ} diagram
at different redshift bins for the simulated and survey samples,
finding they follow similar trends in this plane: as redshift
increases, the proportion of objects neighbouring the dusty
star-forming region decreases, with the distributions becoming
more localized in the star-forming region. However, the
SED-inferred colours of the TNG50 sample peak at different
positions than observations, being more notorious in the \( V - J
\) direction. In \cref{fig:uvj_tng50_vs_jwst_by_mass}, we focus on
the \( 3 \leqslant z < 3.5 \) range and divide the sample by true
stellar mass, finding that, although the synthetic sample is in
general compatible with the expected colours from JWST
observations, the colour discrepancy is present in all mass bins:
for \( \log_{10} \left(\mast / \unit{\msun}\right) > 10\) the
sample is slightly bluer in the \( V - J \) direction than
observations, and redder at lower masses.

In relation to this, \citet{Gebek2025} recently studied the mass
dependence of the \emph{UVJ} diagram at \( z = 2 \) in TNG100 and
found similar trends to those we find here, but with larger
offsets at \( \log_{10} \left(\mast / \unit{\msun}\right) < 10.5
\) and significantly bluer colours at the high-mass end of their
simulation sample compared to JWST observations. We note that,
although the methods used here and in \citet{Gebek2025} are
similar, their colour estimates are based on direct fluxes from
radiative transfer simulations, while ours rely on SED fitting.
Although this could be a significant source of discrepancy, the
trends we find are consistent with their results, suggesting a
common origin for the colour offset. Here we have set a
conservative value of \( f_\text{dust} = 0.3 \), consistent with
previous studies at lower redshifts
\citep[e.g.][]{Camps2016,Rodriguez-Gomez2019}, though this value
may not be optimal at higher redshifts. In this regard, lowering
this ratio could shift synthetic colours blueward, but may bring
tensions at the high-mass end, where dust-reddened galaxies can
appear (which is hinted in the far-right panel of
\cref{fig:uvj_tng50_vs_jwst_by_mass}).

Theoretical and observational studies
\cite[e.g.][]{Cia2013,Vogelsberger2020} suggest that the
dust-to-metal ratio may evolve with redshift. For instance,
\citet{Vogelsberger2020} used the full IllustrisTNG suite to
calibrate this parameter against observed rest-frame UV luminosity
functions, finding that a power-law relation with redshift
provides an acceptable fit to the data. Coincidentally, the values
suggested by this relation at \( z \sim 3 \) are consistent with
the ones used in this work. {{Alternatively, a stellar
mass-dependent dust-to-metal ratio could also help to alleviate
the colour discrepancy, as more massive galaxies are expected to
be dustier, while low-mass systems likely have lower dust
fractions \citep[see e.g.][]{Pannella2015,McLure2018,Shivaei2020}.
This would imply that the colours of low-mass galaxies become
bluer, whereas those of high-mass systems remain similar or become
redder}}. A more plausible explanation for the offsets is the lack
of a proper dust model in the simulation that fully accounts for
the star-dust geometry. For instance, \citet{Gebek2025} determined
that screen-like dust configurations are efficient in matching the
colours of massive high-redshift galaxies with observations.
Similarly, \citet{Akins2022} showed that a dust model coupled with
a galaxy formation model can reproduce the general trends of the
high-redshift \emph{UVJ} diagram but fail to populate the dusty
star-forming region of the diagram with high-mass galaxies. This
highlights the challenges that pose the inclusion of a component
like dust, whose real abundance and distribution within
high-redshift galaxies is still not fully determined, and also
highlights the need of explicit dust models into galaxy formation
simulations that can aid to overcome these challenges.

\section{Conclusions}
    \label{sec:conclusions}

In this work, we have used the TNG50 cosmological simulation
to generate over \num{10000} NIRCam-like synthetic
observations of galaxies at \( 3 \leqslant z \leqslant 6 \),
designed to match the depth and resolution of deep JWST
surveys. Our methodology involves post-processing the
simulation outputs using radiative transfer, accounting for
dust attenuation and scattering within the galaxies, and
subsequently generating synthetic photometry from the
resulting images.

As an application of these synthetic observations, we have
analysed the performance of SED fitting on the simulated sample
and compared the results with JWST observations from the NGDEEP
and JADES surveys. Our main findings are summarized as follows:

\begin{itemize}[leftmargin=*]
    \item At \( z \leqslant 5 \), we {{generally}} recovered
    photometric redshifts with a mean bias close to zero and a low
    outlier fraction; {{at \( z = 4 \) and \( 6 \), the outlier
    fraction is the highest, at \qty{12}{\percentt} and
    \qty{54}{\percentt}, respectively.}}
    \item Redshift residuals show a slight trend with stellar
    mass, with the largest discrepancies found at the low-mass
    end.
    \item For true stellar masses \( \log_{10} \left(\mast /
    \unit{\msun}\right) \leqslant 10 \), SED-inferred masses are
    well recovered within \qty{0.3}{\dex}.
    \item Despite this, we find a systematic underestimation
    of stellar masses, with a flat mean offset of at most
    \qty{-0.23}{\dex}, which becomes more significant at higher masses
    and persists across all redshifts.
    \item The SED-inferred \emph{UVJ} diagram of the TNG50
    sample at \( z = 3 \) is dominated by star-forming
    galaxies, with no objects classified as quiescent
    according to commonly used criteria.
    \item These colours show correlations with several galaxy
    properties such as specific star formation rate, stellar
    mass, mass-weighted age, \emph{V}-band attenuation, and
    metallicity that are in agreement with expected trends at
    low and intermediate redshifts.
    \item By direct comparison of observer-frame NIRCam colours
    and magnitudes, we find that the TNG50 sample broadly follows
    the median colour trends of the  NGDEEP and JADES
    distributions, but is offset towards redder colours at fixed
    absolute magnitudes, across the redshift range considered.
    \item Applying a colour-based definition for quiescence, we
    find that the TNG50 mass-dependent quiescent fraction is in
    reasonable agreement with observations up to \( z \sim 5 \):
    the fraction rises non-monotonically with stellar mass up to
    \( \log_{10} \left(\mast / \unit{\msun}\right) \sim
    \numrange[range-phrase=\text{~--~}]{10.25}{10.75} \), then
    flattens or declines at higher masses.
    \item The best-fit \emph{UVJ} colours of the TNG50 sample are
    compatible with the expected colours of real JWST galaxies,
    but show a primary offset towards redder \( V - J \) colours
    regardless of redshift.
    \item At \( 3 \leqslant z < 3.5 \), we find that this
    discrepancy is present regardless of mass, with the TNG50
    sample being redder at intermediate and low masses.
\end{itemize}

\section*{Acknowledgements}
AGO thanks CONAHCyT for a PhD scholarship. AGO and GB acknowledge
financial support from Universidad Nacional Autónoma de México
through grants DGAPA/PAPIIT IG100319, BG100622 and IN106124. This
work is based on observations made with the NASA/ESA/CSA {\it
JWST}. The data were obtained from the Mikulski Archive for Space
Telescopes at the Space Telescope Science Institute, which is
operated by the Association of Universities for Research in
Astronomy, Inc., under NASA contract NAS 5-03127 for {\it JWST}.
These observations are associated with NGDEEP (program \#2079) and
JADES (programs \#1180, \#1181, \#1210 and \#1286). The
IllustrisTNG flagship simulations were run on the HazelHen Cray
XC40 supercomputer at the High Performance Computing Center
Stuttgart (HLRS) as part of project GCS-ILLU of the Gauss Centre
for Supercomputing (GCS). Ancillary and test runs of the project
were also run on the compute cluster operated by HITS, on the
Stampede supercomputer at TACC/XSEDE (allocation AST140063), at
the Hydra and Draco supercomputers at the Max Planck Computing and
Data Facility, and on the MIT/Harvard computing facilities
supported by FAS and MIT MKI.

\section*{Data availability}
The data from the IllustrisTNG simulation used in this work are
publicly available at the website
\href{https://www.tng-project.org}{https://www.tng-project.org}
\citep{Nelson2019}. The NGDEEP and JADES observations can be
accessed via \citet[]{Leung2023} and \citet[]{Rieke2023},
respectively.

\bibliographystyle{mnras}
\bibliography{bib.bib}

\appendix
\section{}
    \label{sec:appendix}

\subsection{Photometric redshift estimates with new templates}
\label{subsec:appendix_photz}

\begin{figure}
    \includegraphics{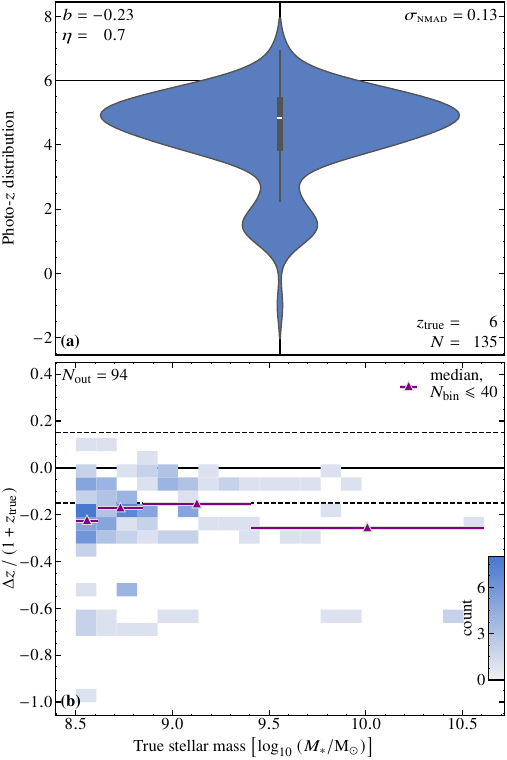}
    \caption{Photometric redshift estimates for the TNG50 sample
    at \( z = 6 \) using the new template set from
    \citet{Larson2023}. The distribution shape remains similar to
    \cref{fig:photzvstruez}, with a mean bias of \( -0.23 \), an
    NMAD of \qty{0.12}{dex}, and an outlier fraction of
    \qty{70}{\percentt}. While the number of low-redshift
    solutions at \( z \sim 1 \) decreases, the overall outlier
    fraction increases due to a shift of the distribution toward
    lower redshifts.}
    \label{fig:appendix_photz}
\end{figure}

We tested the redshift estimation for the \( z = 6 \) sample
using the template set recommended by \citet{Larson2023} for
\( z \sim 4 - 7 \). While this reduced the number of
low-redshift solutions at \( z \sim 1 \), the overall
performance metrics showed no significant improvement. In
\cref{fig:appendix_photz}, we present the redshift recovery
results using the new template set. The distribution shape
remains similar to that in \cref{fig:photzvstruez}, but the
main peak is shifted to lower redshifts. Compared to the
fiducial run, the mean bias remained at \( -0.23 \), the NMAD
decreased slightly to \qty{0.13}{dex}, and the outlier
fraction increased to \qty{70}{\percentt}.

\subsection{Stellar mass recovery with different priors}
\label{subsec:appendix_masses}

\begin{figure*}
    \includegraphics{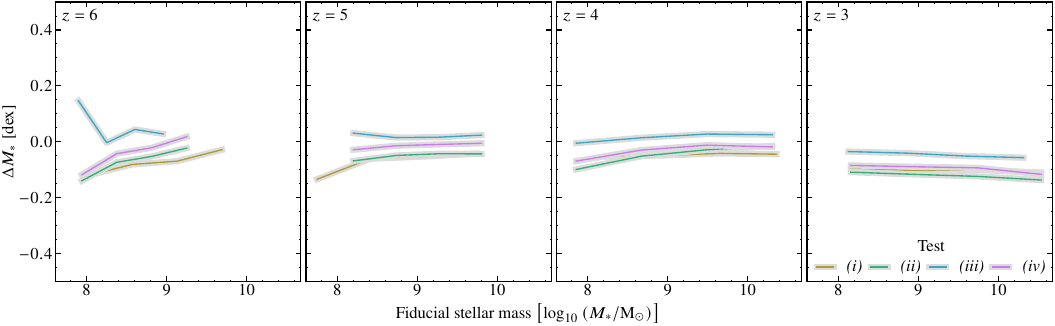}
    \caption{Mass residuals {{(relative to the fiducial estimate)
    plotted against}} fiducial stellar mass for the TNG50 sample
    at \( z = 6, 5, 4, 3 \) using different combinations of priors
    for the dust model and SFH. Overall, we mostly find flat
    offsets of at most \qty{0.15}{dex}, indicating that the choice
    of priors does not abruptly affect the overall trends and that
    the fiducial setup is consistent with alternative prior
    configurations. The legend numbers correspond to the following
    configurations in increasing order: a double power-law SFH
    with a \citet{Salim2018} dust model, a double power-law SFH
    with a \citet{Calzetti2000} dust model, a lognormal SFH with a
    \citet{Calzetti2000} dust model, and an exponential SFH with a
    \citet{Calzetti2000} dust model. The fiducial run combines a
    lognormal SFH with a \citet{Salim2018} dust model.}
    \label{fig:appendix_masses}
\end{figure*}

We repeated the fitting procedure with BAGPIPES, varying the
priors for the dust model and SFH while keeping the remaining
parameters fixed. The additional prior combinations considered
were:
\begin{enumerate*}[leftmargin=*, label=\emph{(\roman*)}]
    \item a double power-law SFH and a \citet{Salim2018} dust model,
    \item a double power-law SFH and a \citet{Calzetti2000} dust model,
    \item a lognormal SFH and a \citet{Calzetti2000} dust
    model, and
    \item an exponential SFH and a \citet{Calzetti2000} dust
    model.
\end{enumerate*}
Full details about of the priors used for these tests are
given in \cref{tab:sedfittingconfigs}. In
\cref{fig:appendix_masses}, we show the mass residual between
the fiducial and new runs against the fiducial stellar mass
for the subsamples at \( z = 6, 5, 4, 3 \). At \( z \leqslant
5 \), all prior combinations exhibit a flat underestimation of
at most \qty{\sim0.1}{dex}, regardless of fiducial mass. At \(
z = 6 \), the offset for combination (ii) decreases from
\qty{\sim0.15}{dex} at \( \log_{10} \left( \mast /
\unit{\msun} \right) \sim 8 \) to around zero at higher
masses, while the other combinations show the opposite trend.
These test results suggest that the choice of priors for the
dust model and SFH does not significantly affect the overall
trends, and that the fiducial setup is consistent with the
alternative configurations across the redshift range
considered.

\begin{table*}
    \centering
    \caption{Configurations used in the SED fitting tests for the
    TNG50 sample. Each configuration combines a specific star
    formation history (SFH) model and a dust attenuation model.
    The columns list the parameter symbols and their corresponding
    units for each component. Except for the total stellar mass
    formed, which has a logarithmic range, all parameters are
    uniformly sampled within the specified ranges.}
    \label{tab:sedfittingconfigs}
    \sisetup{
            exponent-mode = threshold,
            exponent-thresholds = -4:4,
            range-exponents = individual,
            parse-numbers = false,
            input-decimal-markers={,},
            output-decimal-marker = \text{~to~},
            }
    \begin{tblr}{
        width=\textwidth,
        colspec={X[l] X[c] X[c] Q[si={},c] X[l] X[c] Q[si={},c]},
        row{1} = {belowsep=0pt, abovesep=0pt},
        row{1} = {guard},
        rowsep=0pt,
        row{5,10,15,19} = {belowsep+=1pt},
        cell{12}{7} = {guard},
        cell{17}{7} = {guard},
        cell{21}{7} = {guard},
    }
        \toprule
        Configuration & SFH model & Symbol (unit) & Range & Dust model & Symbol (unit) & Range \\
        \midrule
        \SetCell[r=4]{m} Fiducial & \SetCell[r=4]{m}
        Lognormal & \( \mast \, \left( \unit{\msun} \right) \) & 1,10^{13} & \SetCell[r=4]{m} \citet{Salim2018} & \( \delta \) & -1.6,0.4 \\
        & & \( Z \, \left( \unit{\zsun} \right) \) & 0.001,2.5 & & \( B \) & 0.0,10.0 \\
        & & \( t_\text{max} \, \left( \unit{\giga\year} \right) \) & 0.001,15.0 & & \( A_V \, \left( \unit{\mag} \right) \) & 0.01,8.0 \\
        & & \( \sigma_\text{SFH} \, \left( \unit{\giga\year} \right) \) & 0.001,20.0 & & \( \eta \) & 1.0,3.0 \\
        
        \SetCell[r=5]{m} Test \emph{(i)} & \SetCell[r=5]{m} Double power-law & \( \mast \, \left( \unit{\msun} \right) \) & 1,10^{13} & \SetCell[r=4]{m} \citet{Salim2018} & \( \delta \) & -1.6,0.4 \\
        & & \( Z \, \left( \unit{\zsun} \right) \) & 0.001,2.5 & & \( B \) & 0.0,10.0 \\
        & & \( \alpha \) & 0.001,15.0 & & \( A_V \, \left( \unit{\mag} \right) \) & 0.01,8.0 \\
        & & \( \beta \) & 0.001,20.0 & & \( \eta \) & 1.0,3.0 \\
        & & \( \tau (\unit{\giga\yr}) \) & 0.1,15.0 & & & \\
        
        \SetCell[r=5]{m} Test \emph{(ii)} & \SetCell[r=5]{m} Double power-law & \( \mast \, \left( \unit{\msun} \right) \) & 1,10^{13} & \SetCell[r=2]{m} \citet{Calzetti2000} & \( A_V \, \left( \unit{\mag} \right) \) & 0.01,6.0 \\
        & & \( Z \, \left( \unit{\zsun} \right) \) & 0.001,2.5 & & \( \eta \) & 1.0 \\
        & & \( \alpha \) & 0.001,15.0 & & & \\
        & & \( \beta \) & 0.001,20.0 & & & \\
        & & \( \tau \, \left( \unit{\giga\year} \right) \) & 0.1,15.0 & & & \\               
        
        \SetCell[r=4]{m} Test \emph{(iii)} & \SetCell[r=4]{m} Lognormal & \( \mast \, \left( \unit{\msun} \right) \) & 1,10^{13} & \SetCell[r=2]{m} \citet{Calzetti2000} & \( A_V \, \left( \unit{\mag} \right) \) & 0.01,6.0 \\
        & & \( Z \, \left( \unit{\zsun} \right) \) & 0.001,2.5 & & \( \eta \) & 1.0 \\
        & & \( t_\text{max} \, \left( \unit{\giga\year} \right) \) & 0.001,15.0 & & & \\
        & & \( \sigma_\text{SFH} \, \left( \unit{\giga\year} \right) \) & 0.001,20.0 & & & \\
        
        \SetCell[r=4]{m} Test \emph{(iv)} & \SetCell[r=4]{m} Exponential & \( \mast \, \left( \unit{\msun} \right) \) & 10^{4},10^{13} & \SetCell[r=2]{m} \citet{Calzetti2000} & \( A_V \, \left( \unit{\mag} \right) \) & 0.01,6.0 \\                        
        & & \( Z \, \left( \unit{\zsun} \right) \) & 0.0,2.5 & & \( \eta \) & 1.0 \\
        & & \(T_0 \, \left( \unit{\giga\year} \right) \) & 0.002,13.0 & & & \\
        & & \( \tau \, \left( \unit{\giga\year} \right) \) & 0.1,14.0 & & & \\
        \bottomrule       
    \end{tblr}
\end{table*}

\end{document}